\documentclass[12pt]{article}
\usepackage{amsmath}
\usepackage{graphicx,psfrag,epsf}
\usepackage{enumerate}
\usepackage[round]{natbib}

\usepackage{tkz-graph}  
\usetikzlibrary{positioning}
\usepackage{subcaption}

\usepackage{xcolor}
\usepackage{verbatim}
\usepackage{enumerate}
\usepackage{setspace}
\usepackage{natbib}
\usepackage{amsmath,bm}

\usepackage{authblk}            % Multiple authors

\usepackage{url} % not crucial - just used below for the URL 

\newcommand{\blind}{0}

\graphicspath{{./figures/}}

\definecolor{firebrick}{HTML}{b22222}
\definecolor{burntorange}{HTML}{BF5700}
\definecolor{dodgerblue}{HTML}{005A9C}

\newcommand{\head}[1]{{\noindent\textbf{ #1}}}

\newcommand{\X}{\mathcal{X}}
\newcommand{\N}{\mathcal{N}}
\newcommand{\dee}{\mathrm{d}}

\begin{document}

\def\spacingset#1{\renewcommand{\baselinestretch}%
{#1}\small\normalsize} \spacingset{1}

% -------------------------------------------------------------------------
% Title and authors

% Title for both blinded and unblinded versions...
\newcommand{\mytitle}{Bayesian Model Calibration for Extrapolative
  Prediction via Gibbs~Posteriors}

\if0\blind
{
  \title{\bf \mytitle}
  
  % \author{Author 1\thanks{
  %     The authors gratefully acknowledge \textit{please remember to list all relevant funding sources in the unblinded version}}\hspace{.2cm}\\
  %   Department of YYY, University of XXX\\
  %   and \\
  %   Author 2 \\
  %   Department of ZZZ, University of WWW}
  
%   more compact way

  \author[1]{Spencer Woody\thanks{Corresponding author. Email to
    \texttt{spencer.woody@utexas.edu}}}

\author[1]{Novin Ghaffari}

\author[2]{Lauren Hund}

\affil[1]{The University of Texas at Austin, USA}

\affil[2]{Sandia National Laboratories, Albuquerque, NM, USA}

%   \author[1 2]{Carlos Carvalho}

% \author[1]{Jared Murray}

% \author[2]{Spencer Woody\thanks{Corresponding author. Email to
%     \texttt{spencer.woody@utexas.edu}}}

% \affil[1]{Department of Information, Risk, and Operations Management, University~of~Texas~at~Austin}

% \affil[2]{Department of Statistics and Data Sciences, University~of~Texas~at~Austin}

  \maketitle
} \fi

\if1\blind
{
  \bigskip
  \bigskip
  \bigskip
  \begin{center}
    {\LARGE\bf \mytitle}
    
    % Optionally, if you want the date too

    \bigskip
    \today
    
\end{center}
  \medskip
} \fi

% -------------------------------------------------------------------------
% Abstract and keywords

\bigskip
\begin{abstract}
  The current standard Bayesian approach to model calibration, which
  assigns a Gaussian process prior to the discrepancy term, often
  suffers from issues of unidentifiability and computational
  complexity and instability.  When the goal is to quantify
  uncertainty in physical parameters for extrapolative prediction,
  then there is no need to perform inference on the discrepancy term.
  With this in mind, we introduce Gibbs posteriors as an alternative
  Bayesian method for model calibration, which updates the prior with
  a loss function connecting the data to the parameter.  The target of
  inference is the physical parameter value which minimizes the
  expected loss.  We propose to tune the loss scale of the Gibbs
  posterior to maintain nominal frequentist coverage under assumptions
  of the form of model discrepancy, and present a bootstrap
  implementation for approximating coverage rates.  Our approach is
  highly modular, allowing an analyst to easily encode a wide variety
  of such assumptions.  Furthermore, we provide a principled method of
  combining posteriors calculated from data subsets.  We apply our
  methods to data from an experiment measuring the material
  properties of tantalum.
\end{abstract}

% 3 - 6 keywords
\noindent%
{\it Keywords:} Inverse problem; misspecified models; physical models;
uncertainty quantification \vfill

% -------------------------------------------------------------------------
% Set spacing for main text; don't change this

\newpage
\spacingset{1.45} % DON'T change the spacing!

% -------------------------------------------------------------------------
% Main body

% \doublespacing

\section{Introduction}

Model calibration is the process of learning about model inputs by
coupling experimental data with computational simulation outputs,
essentially solving an inverse problem for a deterministic computer
model.  Typically experimental data are assumed to be generated from the process:
\begin{align}\label{eq:obs}
  \begin{split}
    y(x) &= \zeta(x; \theta^*) + \epsilon(x) \\
    \eta(x; \theta^*) &= \zeta(x; \theta^*) + \delta(x; \theta^*)
  \end{split}
\end{align}
where $y(x)$ are observations of the true physical process
$\zeta(x; \theta^*)$ at time $x$ measured with independent errors
$\epsilon(x)$.  The parameter $\theta$ is an input to the physical
process and $\theta^*$ is the true value of $\theta$ that we are
trying to learn.  The true physical process is approximated by
deterministic physical model $\eta(x; \theta)$, and the goal is to
determine the model inputs $\theta$ from the observations of $y(x)$.
However, there is some systemic bias in this physical model, and so
$\delta(x)$ accounts for this inadequacy and is called the model
discrepancy term.

Model calibration can serve two distinct purposes in computer
modeling: interpolative prediction and extrapolative prediction.  In
interpolative prediction, the model parameters are tuned such that the
model can accurately predict throughout the design space \textit{where
  data have been collected}.  In extrapolative prediction, the
calibrated model results are applied to predict in a space or setting
\textit{where no data have been collected and the model cannot be
  directly validated}.  Extrapolative prediction is common in many
engineering applications, as the computer modeling exercise is often
intended to compensate for gaps in data.  Examples of extrapolative
prediction include calibrating subsystem models to roll-up results
into overall system predictions \citep{li2016role}, using calibration
to estimate parameters with scientific meaning for use in other
applications or models \citep{brown2018,brynjarsdottir2014learning},
and using models to predict to spaces where data cannot be collected
\citep{ling2014selection}.

Physical models are inherently approximations of reality and typically
contain error; thus, model calibration procedures must accommodate for
model discrepancy (called model misspecification in the statistics
literature) to obtain valid estimates and uncertainties on calibrated
input parameters.  The ontological question of whether a ``true''
value of a model input exists in reality is debatable, because there
is no ground truth in a computer simulation.  Here we take the
pragmatic view that a true calibration parameter exists insofar as it
informs extrapolative prediction.  Therefore, we refer to such
calibration parameters as ``physical parameters,'' following
\cite{brynjarsdottir2014learning}.

In this paper we present a Bayesian approach to model calibration
based on Gibbs posteriors.  With this method, the target of inference
is the physical parameter which minimizes a user-defined loss function
in expectation, and we accommodate the use of prior information when
available.  In the remainder of this section, we preview our method as
a solution to issues of identifiability and computational stability
associated with the current standard Bayesian approach, and connect it
with existing ideas in the literature which cast calibration as an
optimization problem.  In Section~\ref{sec:choosew}, we explain our
method in detail and motivate our approach of setting the loss scale
for the Gibbs posterior, of which we present an implementation in
Section~\ref{sec:parametric-bootstrap}.  Then, in Section
\ref{sec:ensemble-calibration} we give a statistically principled
method for combining posteriors calculated from data subsets.
Section~\ref{sec:dataapp} applies our methods to data from an
experiment measuring the dynamic material properties of tantalum.  We
close with a discussion in Section~\ref{sec:discussion}.  The online
supplement contains simulation results, algorithmic details, and
additional figures.

\subsection{Standard Bayesian approach}

Calibration is often a poorly identified problem, where multiple
inputs produce equally valid solutions.  Bayesian inference has been
used extensively in model calibration to avoid optimization challenges
under poorly identified parameters. The most common Bayesian approach
comes from the seminal paper by \cite{kennedy2001bayesian} (hereafter,
KOH), whereby the model in Eq.~\eqref{eq:obs} is completed by
assigning priors to the unknown components:
\begin{align}
  \label{eq:KOH}
  \begin{split}
      y(x) = \eta(x; \theta)\, + &\, \delta(x) + \epsilon(x) \\
    % \theta &\sim \pi(\theta) \\
    % \delta &\sim \text{GP}(0, k(\cdot, \cdot))\\
    % \epsilon(x) &\sim \N(0, \sigma^2)
      \theta \sim \pi(\theta), \quad
      \delta \sim \text{GP}(0, \,&k(\cdot, \cdot)), \quad
      \epsilon(x) \sim \N(0, \sigma^2)
  \end{split}
\end{align}
where $k(\cdot, \cdot)$ denotes the covariance kernel of the Gaussian
process (GP) prior for the discrepancy $\delta$.  This model
accommodates different sources of uncertainty, including input
parameter uncertainty, measurement uncertainty in the experimental
data, and model-form error.  This paper is primarily concerned with
this latter uncertainty, model-form error.

The KOH model allows for model misspecification through incorporation
of the GP prior for the discrepancy.  However, it is well known that
including this term to account for model discrepancy is insufficient,
because $\theta$ and $\delta$ are not jointly identifiable
\citep{loeppky2006computer}; various configurations of the two can
explain the data equally well while having similar density under the
prior. The calibration parameters $\theta$ are only guaranteed to
converge to their true physical values when there is no discrepancy,
or when the discrepancy is mean 0 over $x$ and independent of
$\theta$.  Calibration parameters should not be expected to have
physical interpretations under the KOH model in the presence of model
discrepancy \citep{arendt2012improving, brynjarsdottir2014learning}.
The non-identifiability of the discrepancy function is similar to the
problem of ``spatial confounding'' in the spatial statistics
literature \citep{reich2006effects, hodges2010adding,
  paciorek2010importance}, as it is generally not possible to learn a
systematic, unobserved model bias from the observed data alone.

When model inputs have physical interpretations, model discrepancy
must be carefully considered for valid physical parameter estimation
and uncertainty quantification.  For instance,
\cite{brynjarsdottir2014learning} illustrate that the KOH model
results in biased parameter estimates in the presence of systematic
model discrepancy, and the only way to reduce this bias is to know
\textit{a priori} the form of the discrepancy.  Challenges associated
with physical parameter calibration under model misspecification are
well-documented \citep{arendt2012improving, arendt2012quantification,
  brynjarsdottir2014learning, arendt2016preposterior,
  tuo2016theoretical}, but few solutions exist beyond the standard KOH
treatment of model discrepancy.  More recent work shows that KOH
calibration is not $\ell_2$-consistent and calibration parameter
inferences depend on the model discrepancy prior
\citep{tuo2016theoretical}.  To address this issue,
\cite{plumlee2017bayesian} proposes using a prior on the discrepancy
that is orthogonal to the computer model gradient to produce
calibration results that are consistent with the desired loss
function, similar to projected kernel calibration proposed by
\cite{tuo2017adjustments}.

A common thread across all calibration methodologies is
that the analyst must make assumptions about the form of the model
discrepancy in order to obtain accurate calibration parameter
estimates.  For instance, the KOH model assumes that model discrepancy
can be represented as a Gaussian process; further, inferences about
calibration parameters under KOH are generally only correct when the
Gaussian procss is mean 0 and independent of the model inputs.
\cite{brynjarsdottir2014learning} place stronger prior information on
the Gaussian process under a non-mean 0 discrepancy to improve
inferences based on knowledge of the form of the discrepancy.

\subsection{An alternative Bayesian approach}

The goal of this paper is to introduce a Bayesian approach for
accomodating model discrepancy as an alternative to KOH, namely Gibbs
posteriors. With this method, we specify a prior $\pi(\theta)$ for the
calibration parameters, and provide an update for these parameters by
a loss function connecting the data and the unknown parameters. The
Gibbs posterior is
\begin{align*}
  p_w(\theta \mid y) &\propto \exp(-w l(y, \theta)) \pi(\theta)
\end{align*}
where $l(\cdot, \cdot)$ is a loss function and $w$ is a scalar
apportioning weight between the data and the prior, which we call the
loss scale.  Crucially, as explained in greater detail in
Section~\ref{sec:choosew}, the target of inference here is different
from that under the KOH model; our motivation is to quantify
uncertainty in the $\theta$ value that minimizes the loss
$l(y, \theta)$ in expectation under the data generating process for
$y$.  In this sense, the target has a well-grounded physical
interpretation.  This approach provides a flexible alternative
modeling framework to KOH that allows the analyst to utilize prior
knowledge of how the model is wrong when possible.

The Gibbs posterior framework as introduced here for application to
Bayesian model calibration has several advantages: (i) the target of
inference is as the minimizer of the expected loss, avoiding issues
with the ontological meaning of $\theta$ in KOH, and (ii) we bypass
issues of computational stability associated with estimating the
discrepancy function in the KOH model.  Further, this approach
emphasizes and responds to the need for distinct treatments of
calibration for extrapolative versus interpolative prediction,
analogous to predictive versus explanatory modeling in statistics
\citep{shmueli2010explain}.  For extrapolative prediction,
alternatives to KOH are lacking but needed.  Previous work has
employed power likelihood techniques, a special case the Gibbs
posterior, for calibration \citep{jackson2004efficient,
  mosbach2014bayesian, brown2018}, but to our knowledge none of this
work links power likelihood methods to the emerging statistics
literature on Gibbs posteriors, and it remains an open question how to
select the loss scale $w$. In this paper, we generalize previously
proposed Gibbs posterior techniques for the calibration setting, and
give a method of tuning the choice of loss scale.

\subsection{Calibration as optimization}

% For our purposes, the difference between $\gamma$ and $\theta$ is that 
% $\theta$ is used for extrapolative prediction, whereas $\gamma$ is not
% (i.e., $\gamma$ is a nuisance parameter).
		
Given that $\theta$ is difficult to interpret in the KOH model due to
lack of $\ell_2$ consistency, a recent trend in the calibration
literature is to view calibration as an optimization problem
\citep{tuo2015efficient, wong2017frequentist, gramacy2015calibrating}.
The ``ideal'' value of the physical parameters $\theta$ is defined to
be the minimizer of $\ell_2$ distance between the truth physical
process and the observed data:
\begin{align}
  \label{eq:optim-theta-zeta}
  \theta^\star &= \arg \min_{\theta \in \Theta}
                 \int_x 
                 \left[
                y(x) - \zeta(x; \theta)
                 \right]^2  \dee x.
\end{align}
%where $G(x)$ is the sampling distribution of $x$ values under the
%experimental design, and $w(x)$ is a weighting scheme for the
%observations.  
However, because the true physical process $\zeta(x)$ is unknown,
\cite{wong2017frequentist} propose the estimator minimizing the
$\ell_2$ loss between the computer prediction and the observed data:
\begin{align}
  \label{eq:point-wong}
  \begin{split}
      \hat\theta &= \arg \min_{\theta \in \Theta} \int 
               \left[
               y(x) - \eta(x; \theta)
             \right]^2 \dee G(x)
  \end{split}
\end{align}
where $G(x)$ characterizes a weighting scheme for $x$.  The sampling
distribution of this estimator is approximated with the bootstrap.

While viewing calibration as optimization, as given above, simplifies
the definition of the ``best estimate'' of $\theta$ relative to KOH,
this approach is not satisfactory from a Bayesian perspective.  We
would prefer to give a posterior distribution for $\theta$ conditional
on the observed data, while incorporating prior information on both
the discrepancy and the model parameters when possible.  Our Gibbs
posterior method does exactly this.  We reformulate the inferential
goal of the Bayesian approach as one which quantifies uncertainty on a
$\theta$ value which minimizes an expected loss function, in contrast
to the KOH model which expresses uncertainty in a set of parameters
$\theta$ which are involved in a specified probabilistic model given
in Eqs. \eqref{eq:obs} and \eqref{eq:KOH}.

%%% Local Variables:
%%% mode: latex
%%% TeX-master: "../Draft"
%%% End:

\section{Gibbs posteriors for Bayesian model calibration}
\label{sec:choosew}

When physical parameter estimation is the goal of calibration, then we
have no need to make posterior inference on a discrepancy function
(which only improves interpolative prediction).  With this in mind, we
propose a Gibbs posterior approach to calibration under model
discrepancy.

We continue to assume the data generating mechanism of Eq.~\eqref{eq:obs}. In
connection with the frequentist literature viewing model calibration
as an optimization problem, our method has the inferential goal in
quantifying uncertainty around an unknown parameter which minimizes
the expected value of the loss function $l(y, \theta) $, i.e. the
target is
\begin{align}
  \label{eq:true-min}
  \theta_0 \equiv \theta(F_0) &:= \arg \min_{\theta' \in \Theta}
                                \int l(y, \theta') \dee F_0
\end{align}
with the expectation over $F_0$, the true, unknown data generating
process for $y$, a combination of the unknown physical process
$\zeta(x)$ and measurement error $\epsilon(x)$.

% Key questions... but I think these are addressed adequately.
% \sw{Key question:
%   what is the stochastic part of $y \sim F_0$ in this equation?  To
%   me, it seems like for the observations
%   $y(x) = \eta(x; \theta_0) + \delta(x) + \epsilon(x)$, the ``true''
%   parameters $\theta_0$ and the discrepancy $\delta(x)$ are both
%   \emph{fixed} but unknown, and $\epsilon$ is the only stochastic
%   part.  Then, later on when we set the prior $F^\star_0$, we're
%   really just setting a prior on $\delta$.  Whatever it is, I think we
%   need to be really careful in explaining this} \lh{Note that, when
%   the data generating mechanism in \ref{eq:obs} is known, then the
%   only source of stochasticity in $F_0$ is measurement noise
%   $\epsilon(x)$, because the discrepancy can be considered known.}
    
Given a prior belief distribution on $\pi(\theta)$ for the minimizer
$\theta_0$ defined in Eq. \eqref{eq:true-min} and the observed data
$y(x)$, we propose a posterior update for $\theta_0$ in the form of
\begin{align}
  \label{eq:gibbs-posterior}
  p_w(\theta \mid y) &\propto \exp(-w l(y, \theta)) \pi(\theta)
\end{align}
assuming this admits a proper density, for some $w \geq 0$, called the
\emph{loss scale}, which controls relative weighting of the prior and
the data in the update.  \cite{bissiri2016general} show that
Eq. \eqref{eq:gibbs-posterior} constitutes a ``valid and coherent'' update
from the prior $\pi(\theta)$ concerning $\theta_0$, in that, for
example, it follows a sequential invariance property of updating, and
minimizes an additive loss measuring respective discrepancy between
the posterior and prior, and the posterior and data.  The update in
the form of Eq.~\eqref{eq:gibbs-posterior} is referred to in the
literature as the Gibbs posterior \citep{zhang2006, jiang2008,
  alquier2016properties}.  We do not incorporate the discrepancy into
the posterior inference scheme, though we do use prior knowledge on it
to inform the choice of loss scale, as we explain in the following
subsection.

A special case of the Gibbs posterior is power-likelihood methods
\citep{miller2015robust, grunwald2017inconsistency}, where the
likelihood for the (presumably misspecified) probability model is
scaled by some power.  However, here we are more generally interested
in linking the observed data and unknown parameters via a loss
function.  Note that \eqref{eq:gibbs-posterior} accommodates
power-likelihood posteriors as a special case when $l(\cdot, \cdot)$
is the negative log-likelihood, and further, the ordinary Bayesian
posterior is returned when $w = 1$.

The choice of loss function defines the estimand, as shown in
Eq.~\eqref{eq:true-min}.  To choose the loss function, we can
consider prior knowledge about model discrepancy.  For instance, one
loss function could be the $\ell_2$ distance between the observed data
and the expected value of the true process:
\begin{eqnarray}
l(y, \theta) = \int [y(x) - \text{E}_{F_0}(\zeta(x; \theta))]^2 \dee x, \label{goodloss}
\end{eqnarray}
which admits an estimand similar to that in
Eq.~\eqref{eq:optim-theta-zeta}.  When $\text{E}_{F_0}(\delta(x)) = 0$
and so $\text{E}_{F_0}(\zeta(x; \theta)) = \eta(x;\theta)$, this loss
function is the same as that used in the frequentist estimator in
Eq.~\eqref{eq:point-wong} when using a uniform weighting scheme on the
sample space $x \in \X$.

The main benefit of using the Gibbs posterior is now we have a
well-defined target of inference coming from a user-defined loss
function, in contrast to the KOH approach which has a less concrete
target.  As we accumulate more data, we concentrate on the empirical
risk minimizer, as long as this value has positive support under the
prior.  Also, we can avoid the computational difficulties typical in
taking the KOH approach (i.e., calculating matrix inversions in the
MCMC scheme necessary when using a GP prior).  We now address the
question of setting the loss scale $w$.

\subsection{Choosing the loss scale by calibrating credible intervals}
\label{sec:selecting-scale}

Gibbs posteriors, in particular power-likelihood methods, have
previously been applied to model calibration
\citep{jackson2004efficient, mosbach2014bayesian, brown2018}, but it
remains an open question as to how the loss scale $w$ should be
selected.  Outside of calibration applications, there are many other
suggested for approaches \citep{miller2015robust,
  grunwald2017inconsistency, jiang2008gibbs, holmes2017assigning,
  syring2017calibrating, lyddon2017generalized}.
% See Appendix A for more details.  \sw{I don't think we need to have
% this appendix since it's not our work.}
So far there has been no universally applicable approach, though it is
clear that any valid approach should require assumptions of the form of
the discrepancy

In ideal situations, there exists an analytic solution.  As an
example, if we are willing to make the strong assumption that the
model discrepancy arises from a mean 0 Gaussian process over $x$ and
apply asymptotic results, then we can calculate the asymptotic
posterior distribution for $\theta$ under the Gibbs posterior
\eqref{eq:gibbs-posterior}.  \cite{brown2018} use this approach to
justify scaling the likelihood by a function of the effective sample
size of the residuals.  However, heavily relying on asymptotic results
may not be prudent, given that many calibration problems involve
sparse data and, more importantly, this approach relies on strong
assumptions about the form of the model discrepancy (mean 0 Gaussian
process).  A primary advantage of our Gibbs posterior approach to
calibration is that this framework allows for a very broad class of
such assumptions.

For the calibration problem, we propose the strategy of tuning the
loss scale $w$ to achieve correct nominal frequentist coverage of the
Gibbs posterior under the assumptions of the discrepancy, as
approximated by a bootstrap approach.  In this way, our method is
similar in spirit to \cite{syring2017calibrating} and
\cite{lyddon2017generalized}.

Given that $F_0$, the data generating process for $y$ in
Eq. \eqref{eq:true-min}, is unknown (because the model discrepancy is
unknown), we may assign it a prior $F_0 \sim F^\star_0$.  This prior
then propagates uncertainty onto the distribution for the minimizer
$\theta_0$ via the hierarchical formulation
\begin{align}
  \label{eq:bootstrap-model}
  \begin{split}
    F &\sim F^\star_0 \\
    (\theta(F) \mid F) &= \arg\min_{\theta' \in \Theta}
    \int l(y, \theta') \dee F.
  \end{split}
\end{align}
We refer to~\eqref{eq:bootstrap-model} as the bootstrap model, since
implementation generally requires sampling over the prior distribution on the
data generating mechanism.  Note that, in the bootstrapped model in
\eqref{eq:bootstrap-model}, the data generating mechanism is
considered a random variable, as opposed to Eq.  \eqref{eq:true-min}
where the data generating mechanism $F_0$ is fixed but unknown.  As a
result, in the bootstrapped model, there is stochasticity in $y(x)$
due to measurement noise and in $\eta(x; \theta)$ due to unknown model
discrepancy.

\cite{lyddon2017generalized} consider a similar approach for the case
of independent observations arising from $F_0$ and select the $w$ for
the Gibbs posterior by matching the Fisher information of the
asymptotic distributions for the Gibbs posterior
\eqref{eq:gibbs-posterior} and bootstrap model
\eqref{eq:bootstrap-model}.  However, deriving similar analytic
results for the calibration problem is likely intractable and perhaps
not the most pragmatic choice in our case.

Instead, we propose selecting the loss scale $w$ to retain frequentist
coverage for the Gibbs posterior, following
\cite{syring2017calibrating}.  We argue that coverage is a natural
metric for the calibration application, since our primary interest is
sufficient uncertainty quantification (UQ) for the physical parameters
of interest.  One reasonable definition of sufficient UQ in this sense
is accurate credible interval width under our prior assumptions on the
model discrepancy.  Such a requirement is stemmed in the property
that, for a correctly specified model credible intervals for
parameters retain frequentist coverage on average across the prior\footnote{Note that coverage of credible
  intervals is not uniform across all $\theta$, or equivalently,
  conditional on any particular $\theta$, with the exception of the
  special class of ``matching priors'' \citep{ghosh2011}, i.e., in
  general
  $$\text{Pr}_{P_\theta}(C_\alpha(\theta \mid y) \ni \theta \mid
  \theta) \neq 1-\alpha. $$}.  Generally, for a model given by
$(y \mid \theta) \sim P_\theta$, $\theta \sim \pi(\theta)$, if
$C_\alpha(\theta \mid y)$ is a $1-\alpha$ ordinary posterior credible
interval, it follows that
$$ \int \text{Pr}_{P_\theta}(C_\alpha(\theta \mid y) \ni \theta)
\pi(\dee\theta) = 1 - \alpha. $$ 

%  \begin{align}\nonumber
%      \int \text{Pr}_{P_\theta}(C_\alpha(\theta \mid y) \ni \theta)
%   \pi(\dee\theta)
%   = 1 - \alpha.
% \end{align}

In our case, we want to ensure that the Gibbs posterior retains
frequentist coverage of the target $\theta$ on average under the
presumed prior for the data generating process.  Let
$C_{\alpha, w}(\theta \mid y, \eta)$ denote the $1 - \alpha$ Gibbs posterior
credible interval with scale $w$, and let the average frequentist
coverage be defined as
\begin{align}
  \label{eq:coverage-def}
  c_{\alpha}(w; F_0^\star) &:= \int \text{Pr}_{F^\star_0}(C_{\alpha,
                          w}(\theta \mid y,\eta) \ni \theta) \pi(\dee \theta).
\end{align}
This is the frequentist probability, on average across the prior, that
the Gibbs posterior credible interval with loss scale $w$ contains the
true $\theta$ under the assumed prior $F^\star_0$.  We want to choose
$w$ such that $c_{\alpha}(w; F_0^\star) = 1 - \alpha$.  Such an
approach is closely related to the ``calibrated Bayes'' school of
thought, whereby frequentist properties are used as criteria for
Bayesian model evaluation \citep[][]{little2006, box1980, rubin1984,
  gelman}.

Again, finding the optimal choice of $w$ to uphold the frequentist
coverage property is analytically intractable in most cases.  Instead,
we propose to use Monte Carlo approximations of coverage to evaluate
possible choices of $w$ using a bootstrap
approach.  % Specifically, we require
% that $w$ is set such that $100 \times (1-\alpha)\%$ of samples from
% $\theta(F)$ fall within the Gibbs posterior credible interval.
% \sw{This previous sentence is wrong, I think.... Also, I don't think
%   we need this definition\ldots} That is, we find the solution to
% \begin{align}
%   \hat{w} &= \arg \min_{w \in [0, 1]}
%             \left[
%             \hat{c}_\alpha(w; F^\star_0) - (1 - \alpha)
%             \right]^2,
% \end{align}
% where $\hat{c}(w; F^\star_0)$ denotes the Monte Carlo estimate of
% coverage. % \lh{what is $\hat{c}$?  i'm not convinced you need
%   % $\hat{c}$ at this point.  why can't we just use $c$?} \sw{I think we
%   % do need it, if only to highlight that it's an estimate.  }
To obtain these estimates of frequentist coverage for the Gibbs
posteriors, we need to specify the prior for $F_0^\star$ and generate
data under this prior.  For this, we propose both a parametric
procedure, which we describe in detail in
Section~\ref{sec:parametric-bootstrap}, and a nonparametric procedure,
described in the supplement.

\section{Parametric bootstrap}
\label{sec:parametric-bootstrap}

Here we describe a parametric bootstrap implementation to
approximating Eq. \eqref{eq:coverage-def}, the average frequentist
coverage of the Gibbs posterior with loss scale $w$.  Note that other
implementations of the parametric bootstrap may be reasonable for
scale parameter selection.  The online supplement describes one such
approach, and presents a simulation study which verifies that this
weight selection procedure ensures nominal frequentist coverage under
assumptions of discrepancy form.

We assume that a loss function $l(y, \theta)$ has been specified to
link the data and parameters of interest $\theta$, for example the
$\ell_2$ distance.  Furthermore, we assume that a prior
$\theta \sim \pi(\theta)$ has been specified for the parameters of
interest, as well as a prior for $F^\star_0$, i.e.,
$\delta \sim \pi(\delta \mid \lambda)$ and
$\epsilon \sim \pi(\epsilon \mid \tau)$ are given, with
$\text{E}(\epsilon) = 0$.  Usually there is a Gaussian process prior
for the discrepancy $\delta$ and a Gaussian prior for the measurement
error $\epsilon$, but here we use a more general notation, allowing
these prior to be governed by some arbitrary hyperparameters $\lambda$
and $\tau$, respectively.  These hyperparameters may be
obtained % ; in practice, they are specified by the analyst,
using either empirical Bayesian estimation, expert judgment, or a
combination of the two.  Section \ref{sec:empir-bayes-prior} gives an
empirical Bayesian procedure for estimating the hyperparameters in
order to approximate $F^\star_0$.  We require all priors to be proper
so that samples may be drawn from them.  The bootstrap procedure is
then described as follows:
\begin{enumerate}[(i)]
\item Create bootstrap sample data.  For $b=1,\ldots,B$, draw
  components from the assigned priors for the unknown values,
  \begin{align*}
    \theta^{(b)} \sim \pi(\theta), \quad
    \delta^{(b)} \sim \pi(\delta \mid {\lambda}) , \quad
    \epsilon^{(b)} \sim \pi(\epsilon \mid \tau),
  \end{align*}
  and from these compute the bootstrap observations to mimic draws
  from $F^\star_0$,
  \begin{align*}
    y^{(b)}(x)&= \eta(x; \theta^{(b)}) + \delta^{(b)}(x) + \epsilon^{(b)}(x).
  \end{align*}
\item Given one value of $w$, construct credible intervals
  $C_{\alpha, w}(\theta \mid y^{(b)})$ for each dataset
  $b=1,\ldots,B$.  Then the frequentist coverage is estimated by
  \begin{align*}
    \hat{c}_\alpha(w; F^\star_0)
    &= B^{-1}\sum_{b=1}^B
      \mathbf{1}(\theta^{(b)} \in C_{\alpha, w}(\theta \mid y^{(b)}))
  \end{align*}
\item Choose the loss scale $\hat w$ such
  $\hat{c}_\alpha(\hat w; F^\star_0) \approx 1 - \alpha$, by using a
  defined grid of values or by using a stochastic approximation
  \citep{syring2017calibrating}. Using this, we can form the
  calibrated Gibbs posterior $p_{\hat w}(\theta \mid y)$ using the
  experimental data to perform final inference for $\theta$.
\end{enumerate}

Intuitively, one can see how the prior for $\delta$ (either specified
with expert knowledge or through empirical Bayes) informs the final
choice of loss scale.  If $\delta$ is expected to be erratic and large
in magnitude, then the point estimate for each bootstrap sample
$\hat \theta^{(b)}:=\arg\min_\theta l(y^{(b)}, \theta)$ will be
inaccurate, and therefore the scale $w$ must be smaller (closer to 0)
so that the Gibbs posterior is more diffuse to maintain the right
coverage.  Conversely, if $\delta$ is expected to be smooth and small
in magnitude, then $\hat \theta^{(b)}$ will be close to the truth
$\theta^{(b)}$, and $w$ will be tuned to be larger.

Note that the necessary prior specifications are similarly required
for the standard Bayesian approach of KOH.  The difference here is
that the priors are used here for the purposes of tuning the choice of
$w$.  Unlike under the KOH approach, here the prior $F^\star_0$ is not
something that is updated in the Bayesian procedure, but rather a
modeling assumption which drives the final choice of $w$, and
subsequently influences inference on $\theta$.  In KOH, inferences are
not generally as sensitive to choice of the prior, since the prior is
updated in the posterior sampling scheme.  Hence, using this
parametric bootstrap procedure, careful selection of $F^\star_0$ is
required to obtain valid inference.

% \sw{Move this to discussion.  } This procedure shares conceptual
% similarities to the pre-posterior method from
% \cite{arendt2016preposterior} to determine how identifiable a
% calibration parameter is by repeatedly sampling discrepancy functions
% and evaluating variability in the calibration parameters.  It is also
% similar to the bootstrap procedure used by \cite{wong2017frequentist}
% to estimate the sampling distribution of the frequentist point
% estimate \eqref{eq:point-wong}; however, the underlying goal in our
% case is much different, as our bootstrap procedure is designed to
% calibrate the width of credible intervals from the Gibbs posterior.

\subsection{Empirical Bayes prior}
\label{sec:empir-bayes-prior}

We can derive an empirical Bayes variant of the parametric bootstrap
procedure as follows.  First, find the best fitting estimate
$\hat{\theta} = \arg \min_\theta l(y, \theta)$ and calculate the
empirical discrepancy $\hat{\delta} = y(x) - \eta(x; \hat{\theta}).$
% \begin{align*}
%   % \label{eq:theta-point}
%   \hat{\theta} = \arg \min_\theta l(y, \theta), \quad
%   % \label{eq:delta-point}
%   \hat{\delta} = y(x) - \eta(x; \hat{\theta}).
% \end{align*}
Then, obtain estimates for $\hat\lambda$ and $\hat{\tau}$ (e.g., the
hyperparameters for the GP discrepancy, and the residual variance)
using maximum marginal likelihood on $\hat{\delta}$.  This approach is
similar to \cite{brown2018}, who also calculate the empirical
discrepancy, in their case to estimate the effective sample size of
this discrepancy to determine the appropriate scale $w$.  Further, it
is analogous to residual resampling in standard parametric bootstrap
approaches for regression models \citep{efron1979}, in that we
``resample'' the estimated functional discrepancy $\hat{\delta}$ in
the bootstrap procedure we describe.

%%% Local Variables:
%%% mode: latex
%%% TeX-master: "../Draft"
%%% End:

\subsection{Toy example}
\label{sec:toy}

To illustrate the proposed method in a simple setting, we repeat the
example given in \cite{brynjarsdottir2014learning}.  Consider the data
generating mechanism:
\begin{eqnarray}\label{eq:truemod}
  y_i = \frac{\theta x_i}{1+x_i/a} + \epsilon_i, \quad \epsilon_i \sim \N(0, \tau)
\end{eqnarray}
where $\theta = 0.65$, $\tau = 0.01^2$, and $a=20$.  The goal of the
calibration is to estimate $\theta$.  To mimic model discrepancy, the
authors assume the computer model approximation is
\begin{eqnarray}\label{eq:fitmod}
  y_i = \theta x_i + \epsilon_i , \quad \epsilon_i \sim \N(0, \tau),
\end{eqnarray}
with $\tau$ unknown.  We generate $n=60$ observations from the true
data generating mechanism, and then consider 4 models for calibrating
$\theta$: (i) Maximum likelihood using Eq.  \eqref{eq:fitmod},
ignoring discrepancy, (ii) the KOH model, (iii) power likelihood model
following \cite{brown2018}, and (iv) Gibbs posterior model with
parametric bootstrap.

\begin{figure}[ht!]
\centering
\begin{subfigure}[b]{0.3\textwidth}
\includegraphics[width=1\textwidth]{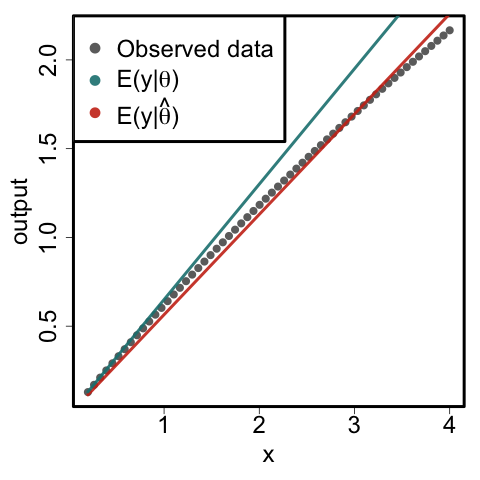}
\subcaption{}
\label{fig:sf1}
\end{subfigure}
\begin{subfigure}[b]{0.3\textwidth}
\includegraphics[width=1\textwidth]{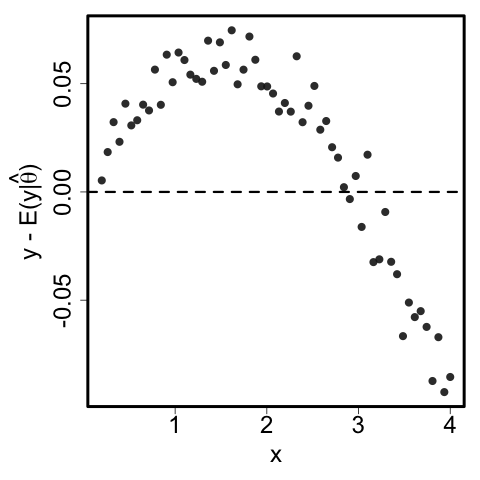}
\subcaption{}
\label{fig:sf2}
\end{subfigure}
\begin{subfigure}[b]{0.3\textwidth}
\includegraphics[width=1\textwidth]{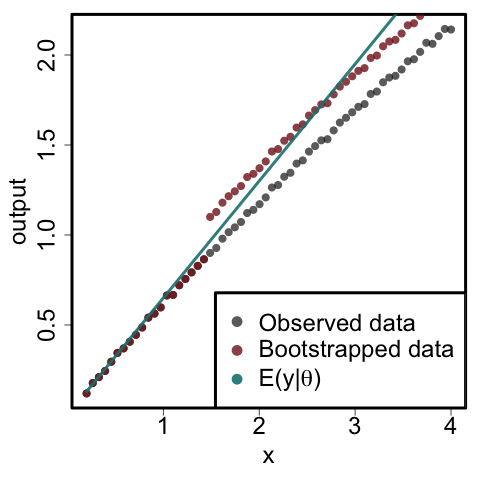}
\subcaption{}
\label{fig:sf3}
\end{subfigure}
\caption{(a) The grey points are the observations sampled from the
  true model; the blue line is the incorrect (linear) model with
  $\theta=0.65$; and the red line is the incorrect (linear) model with
  the ``best fitting'' value of $\theta = 0.57$. (b) The shape of the
  residuals from the best-fitting linear model suggests there is model
  discrepancy. (c) Parametric bootstrapped data (red line)
  approximates the model with discrepancy at the true value of
  $\theta$ (blue line).}
\label{fig:brynexample}
\end{figure}

\paragraph{{Ignoring discrepancy} } First, we fit the model in Eq.
\ref{eq:fitmod} using maximum likelihood, ignoring discrepancy.  The
95\% confidence interval for $\theta$ is $(0.56, 0.57)$, which is far
from the true value.  The true and fitted model are displayed in
Figure \ref{fig:sf1}.

\paragraph{{KOH}} Now, suppose we instead fit a KOH model that allows
for discrepancy:
\begin{eqnarray}
  y_i = \theta x_i + \delta(x_i) + \epsilon_i \label{eq:gpmod}
\end{eqnarray}
We fit the model in Eq.~\eqref{eq:gpmod} using generalized least
squares (GLS) assuming a squared-exponential kernel for the
correlation function.  The 95\% confidence interval for $\theta$ is
(0.48, 0.58), wider than OLS, but still too narrow to encompass the
true value of $\theta$.

\paragraph{{Power-likelihood with effective sample size weighting}}
From the GLS fit, we follow \cite{brown2018} and use the estimated
correlation function parameters to estimate the ESS and find
$n_e \approx 1.4$, i.e. the residuals are highly autocorrelated
(Figure \ref{fig:sf2}).  We now fit a Bayesian linear regression
model, with the log-likelihood scaled by $n_e/n$.  We specify a
disperse normal prior for $\theta$ and an inverse gamma prior for
$\tau$.  Hyperparameters were selected to make these priors proper but
essentially noninformative.  The 95\% credible interval for $\theta$
is (0.43, 0.70) and now brackets the true value of $\theta=0.65$.  The
power-likelihood approach gives exceptionally wide confidence interval
estimates in this case, as we essentially do not have any degrees of
freedom for parameter estimation due to the high residual
autocorrelation.

\paragraph{{Gibbs posterior with bootstrap}} Finally, we implement the
Gibbs posterior method with the parametric bootstrap, specifying a
distribution $F^\star_0$ to encode prior assumptions about model
discrepancy.  The parametric bootstrap procedure in Section
\ref{sec:parametric-bootstrap} allows us to incorporate known
information about discrepancy into the problem.  The model discrepancy
gets worse as $x$ increases, and the magnitude of the discrepancy
ranges between 0 and 0.43.  To mimick this mechanism, we assume a prior
discrepancy of 0 over the first 1/3 of the support of $x$ and a
Uniform(0, 0.4) shift over the latter 2/3 of the support.  Figure
\ref{fig:sf3} illustrates how imposing this prior distribution
corrects for model discrepancy.

%Therefore, we can examine how the
%calibration parameters change as a function of the data included in
%the calibration.  We again calibrate using GLS, but only over subsets
%of the data. We consider 2 subsetting methods: first, we change the
%upper limit of $x$; second, we break the range of $x$ into 4 mutually
%exclusive intervals.  As expected, only calibrating at lower values of
%$x$ results in better inferences, and `throwing out data' is
%advantageous for estimating $\theta$ (Figure \ref{fig:toychunks}).

%\begin{figure} \centering
%\includegraphics[width=.45\textwidth]{Figure_Bryn1.png}
%\includegraphics[width=.45\textwidth]{Figure_Bryn2.png}
%	\caption{Example of `calibrating in chunks.'  (Left) The point
%estimate of $\theta$ moves further from the true value $.65$ as the
%range of $x$ increases to areas where model discrepancy is larger.
%(Right) $\theta$ is estimated using different ranges of $x$, resulting
%in estimates closer to the true $\theta$ in the range of $x$ where
%discrepancy is smallest.}
%	\label{fig:toychunks}
%\end{figure}

With this specfication of $F^\star_0$, we use the $\ell_2$ loss
function in Eq. \eqref{goodloss} and resample $y$ from the
hypothesized DGM, as in Eq. \eqref{eq:obs}.  We obtain a loss scale of
$w \approx 1$, posterior median $\hat{\theta} = -0.64$, and a 95\%
posterior credible interval on $\theta$ of (0.58, 0.70).  We again
bracket the true value of $\theta$, as in the ESS method, but now with
greater precision in the uncertainty interval.  While this example is
contrived, the results clearly illustrate that the parametric
bootstrap can perform well with good prior information on the
discrepancy function.  In the absence of this information, the method
would give poor inference.

\section{Ensemble calibration}
\label{sec:ensemble-calibration}

In some calibration problems, it may be advantageous to divide the
data into subsets, calibrate on the subsets, and then combine
inferences across them.  We consider two primary motivations for
calibrating in subsets: (i) it can be more computationally efficient,
particularly when datasets are large and/or calibration parameters
vary over the model space $x \in \X$; and (ii) when the discrepancy is
non-stationary over $\X$, updating the loss separately for different
subsets of $\X$ will often be simpler than trying to specify a joint
loss and discrepancy prior over all of $\X$.  For example, in the
material model calibration problem described in
Section~\ref{sec:dataapp}, material properties can be calibrated
separately for each experiment and then combined over experiments to
produce a global estimate.  Discrepancy may differ by experiment; by
estimating $\theta$ separately for each one, we do not have to
incorporate the changing magnitude of the discrepancy into the loss
function, and we can gauge how much information each experiment
provides about $\theta$ in the presence of discrepancy.

We refer to the idea of calibrating in subsets as ``ensemble
calibration.''  Section \ref{sec:choosew} considered choosing a single $w$ for the entire dataset. % The coverage calibration methods in Section
% \ref{sec:choosew} selected a single $w$ for the entire dataset.
In ensemble calibration, model misspecification is considered for
subsets of the data.  The data are subdivided over $\X$ into $K$
subsets and each subset receives its own loss scale, so the full
posterior is
 \begin{align}
  p(\theta \mid y) \propto \prod_{k=1}^K
  \exp(-w_k(y_k \mid \theta))\pi(\theta), \label{eq:jointpost}
\end{align}
with the methods in Sections \ref{sec:choosew} and
\ref{sec:parametric-bootstrap} used to select loss scales $w_k$ for the
different subsets.

Directly updating the model in Eq.~\eqref{eq:jointpost} is
disadvantageous.  \cite{gunawan2017bayesian} show that directly
calibrating this joint posterior of the weighted subsets results in
consistent estimates of $\theta$, but uncertainty in $\theta$ is
underestimated. % (they propose a data
% augmentation approach to posterior inference as an alternative to the
% joint posterior.  are calibrated separated for each subset)
An alternative solution is to calibrate each subset separately and
obtain a posterior for each one:
\begin{eqnarray} p(\theta \mid y_k) \propto \exp(-w_k
l(y_k \mid \theta))\pi(\theta), \quad k=1,\ldots,K. \label{eq:obayes2}
\end{eqnarray} 
The posteriors $p(\theta \mid y_k)$ can then be combined \textit{post
  hoc} to produce an overall estimate of the calibration parameters.
This approach has computational advantages, in that calibrating each
subset is simpler than the joint calibration problem.  Further, the
ensemble calibration approach is advantageous because we get model
diagnostics for free.  That is, allowing $\theta$ to vary over $x$
could help diagnose areas of the model space $x$ where model
discrepancy could substantively change the calibration parameters.
Calibrating subsets separately provides information about how
calibration parameter estimates vary across the input space, which
informs how discrepancy impacts the calibration parameter estimates.

Ensemble calibration is closely related to distributed Bayesian
analysis, where data is divided into subsamples for subset inference
\citep{scott2016bayes}. The resulting subset posteriors are combined
into a consensus posterior distribution approximating the full data
posterior. Similarly, ensemble calibration entails solving the
calibration problem on subsets of the data and combining the results.

\subsection{Combining subset posteriors}
\label{sec:comb-subs-post}
We apply the recently developed method of Wasserstein scalable
posteriors (WASP) to combine subset posteriors
\citep{pmlr-v38-srivastava15}. WASP divides the data into $K$ subsets,
finding posteriors $\Pi_k$ for each subset $k=1,\ldots,K$; then,
subset posteriors are combined into a consensus posterior $\bar{\Pi}$
by calculating an approximate Wasserstein barycenter or an ``average''
(with respect to Wasserstein distance) of the subset posterior
distributions.  As long as the number of subsets $K$ is not growing
too quickly with the sample size $n$, WASP achieves almost optimal
convergence to the true parameter, $\theta$, and WASP produces
estimates that asymptotically converge to the full data posterior,
even for many models with independent and non-identically distributed
observations \citep{2015arXiv150805880S}.

Here we use the WASP framework to provide a coherent framework for
combining individually calibrated experiments into a consensus
posterior. For the consensus mean we use an inverse
covariance-weighted average of the subset means. While the Wasserstein
barycenter mean is the unweighted average of individual means, we
think a covariance-weighted average is better here for two
reasons. First in our framework the covariance of each experiment is
not only another parameter but also a quantification of uncertainty
about the parameter of interest. Second, when normal likelihoods are
used, this method provides an exact consensus posterior mean
\cite{scott2016bayes}, if we factor individual loss scales. Finally,
we note that computing the consensus mean in this manner has no effect
on the consensus covariance computation. An iterative algorithm,
described in the supplement, is used to calculate the consensus
covariance matrix.

There is a question of scaling the
consensus posterior. In their original paper, Srivastava et al. use the ``stochastic approximation trick''; each subset likelihood is raised to the power of $K$. This forces the consensus posterior to reflect that we have $nK$ as opposed $n$ total data points. This models the covariance
of parameters \emph{across} all experiments; each experiment provides a noisy estimate of $p(\theta)$ and information across experiments is pooled, providing greater power. Otherwise, omitting the scaling factor $K$, we effectively model
the covariance of parameters $\theta$ \emph{within} an individual experiment, the covariance of an ``average'' experiment. Based on the assumptions about the discrepancy process, $\delta$, we recommend scaling as follows. When discrepancy $\delta$ is not known or assumed to be nonzero mean, then we recommend omitting the $K$ factor rescaling. In this case, the uncertainty from individual experiment posteriors may account for bias that does not disappear as more experiments are conducted. In the case that $\delta$ is known or assumed to be zero mean, has finite variance, and the $x$ for each experiment are coming from a relatively small interval, we may apply the scaling $K$. Here experiments are providing information about parameters of interest for a specific interval of $x$ values, and as the discrepancy is zero mean, we expect parameter estimates to be unbiased. If in addition $\delta$ is stationary, so that its variance is not shifting with $x$, then we may apply the scaling $K$ even to experiments over different values of $x$. Here we expect different experiments, even over different $x$ values, to provide unbiased estimates of parameters $\theta$ with the same scaling of uncertainty.

Combining subset posteriors using the Wasserstein metric has several
advantages.  The Wasserstein metric encodes geometric information
about the distribution and the sample space. For example, the
Wasserstein barycenter between distributions of the same
location-scale family will also be in the same location-scale
family. This does not hold for Euclidean barycenters of distributions
in the same location-scale family.  These factors make the WASP
approach more flexible in the setting of ensemble calibration. There
is no need to down-weight priors, as with other methods, potentially losing conjugacy. The method can also handle
deviations from Gaussianity and different specifications of loss
function.

We outline our algorithmic process for combining subset
posteriors and give additional detail in the online supplement; for
general background in this area, see \cite{Villani2003, villani2006,
  gouicloubes2017, aguehcarlier2011, alvarezestebanetal2016}.

%%% Local Variables:
%%% mode: latex
%%% TeX-master: "../Draft"
%%% End:

% \input{input/Motivating-Example}

\section{Application to dynamic material properties}
\label{sec:dataapp}

Dynamic material models describe how materials behave at extreme
conditions of high pressures and temperatures.  % The experiments
% considered in this example reached peak pressures between 70–240 GPa,
% which are of the order of the pressure in the Earth’s inner core.
To calibrate these material models, we couple computational
predictions of velocity over time with experimental measurements.  In
this application, our goal is to calibrate parameters of the tantalum
equation of state, namely the bulk modulus pressure derivative $B_0'$
based on the model form of \cite{vinet1989}.

\begin{figure}[!ht]
  \centering
\includegraphics[width=.37\textwidth]{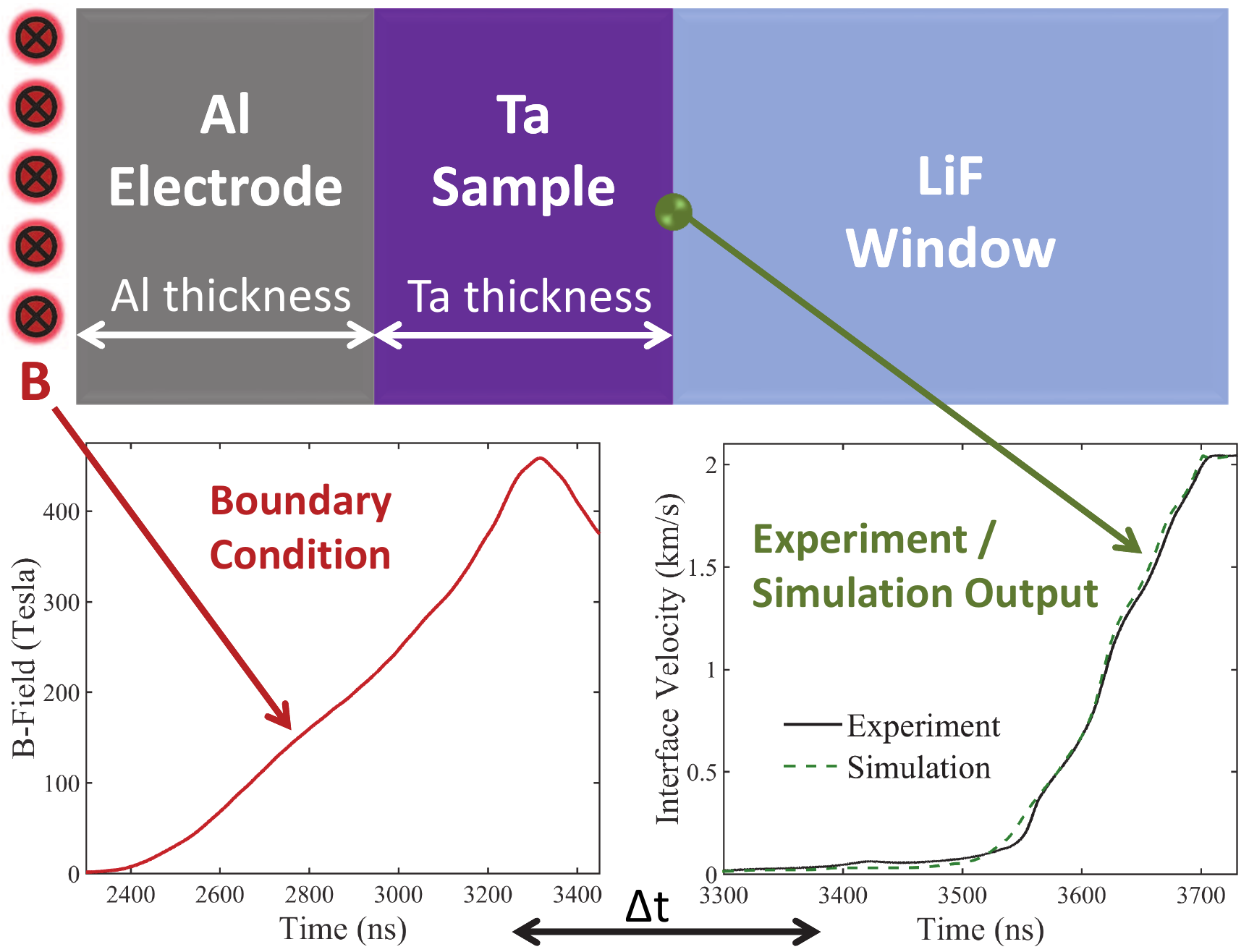}
\includegraphics[width=.5\textwidth]{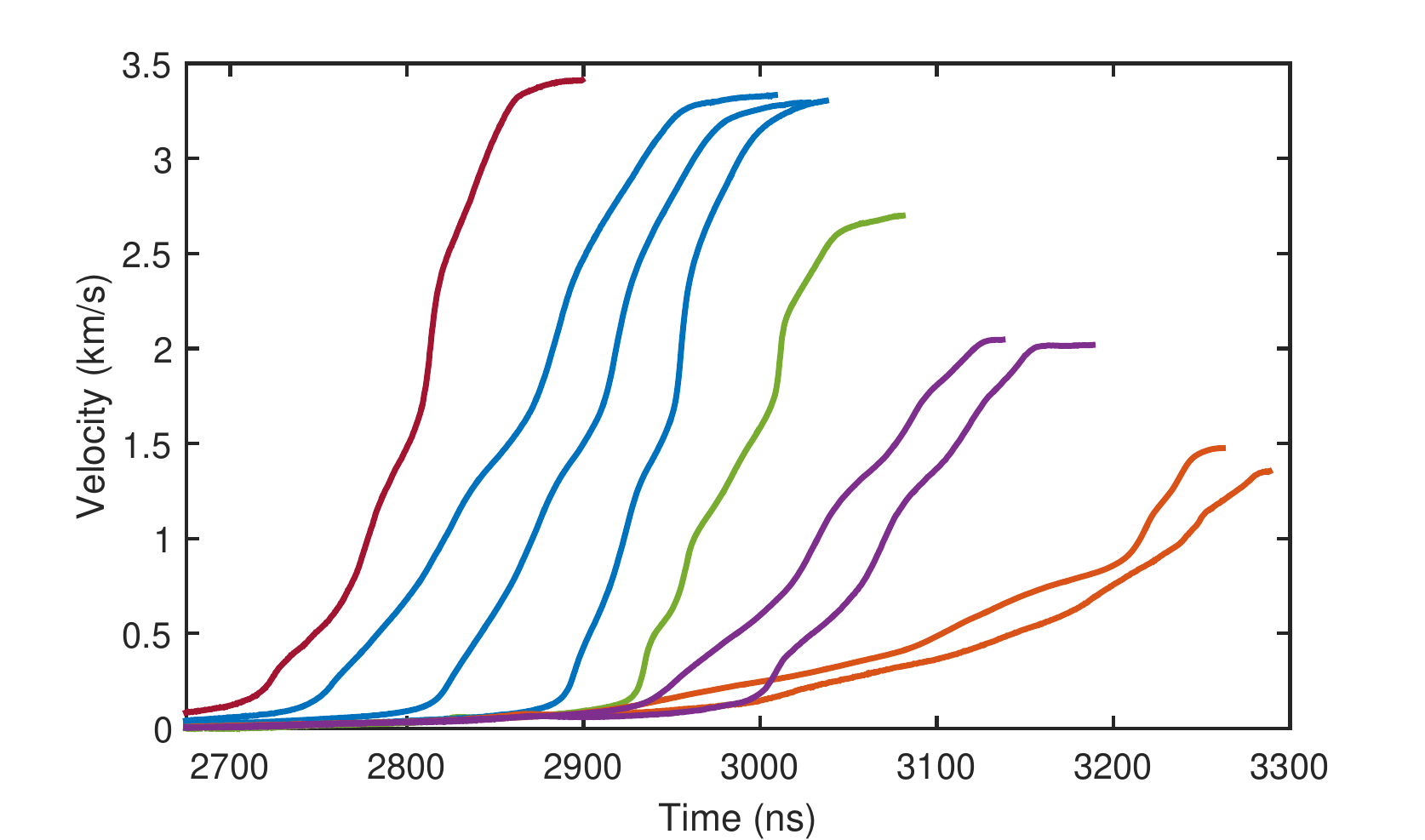}
\caption{\textit{Left}: Experimental configuration.  A time-dependent
  magnetic field (boundary condition) results in a time-dependent
  stress wave propagating through the system; the experimental output
  is the velocity measured at the interface between the tantalum
  sample and lithium-flouride window. \textit{Right}: Nine different
  velocity curves from experiments with different thicknesses and/or
  boundary conditions.}
\label{fig:exp}
\end{figure}

Here we overview the structure of the problem; details are further
discussed in \cite{brown2018}. We analyze the same data here to
illustrate how our method works.  We have data from 9 different
experimental measurements.  In the experiments, a pulsed power driver
delivers massive electrical currents over short time scales through an
aluminum panel, resulting in a time-dependent stress wave (impulse)
propagating through the tantalum samples and then through transparent
lithium fluoride windows (Figure \ref{fig:exp}, left panel).  The
output of interest in the velocity over time at the tantalum-lithium
fluoride interface (Figure \ref{fig:exp}, right
panel).  % The experimentally measured velocities
% were then compared against computational predictions.
Inputs to the computer model predictions include the material
properties of interest for calibration and other experimental
uncertainties, including: uncertainty in the boundary condition, which
specifies the impulse imposed on the material sample; the thicknesses
of the tantalum and aluminum samples; and the initial tantalum
density.

Because the same tantalum plate was used to generate samples across
all experiments, the material properties and density of tantalum do
not vary between experiments.  Differences between the experiments
occur due to different thicknesses of the tantalum samples and
different boundary conditions.  A single experiment produces a
functional output (a velocity curve measured over time) which can be
sampled at a high rate.  While there is a small amount of measurement
uncertainty in velocity at each time point, smoothing this curve can
essentially eliminate this measurement noise at each time point, such
that the experimental outputs can be considered smooth with no
measurement noise.

\begin{figure}[!ht]
\centering
\includegraphics[width=.45\textwidth]{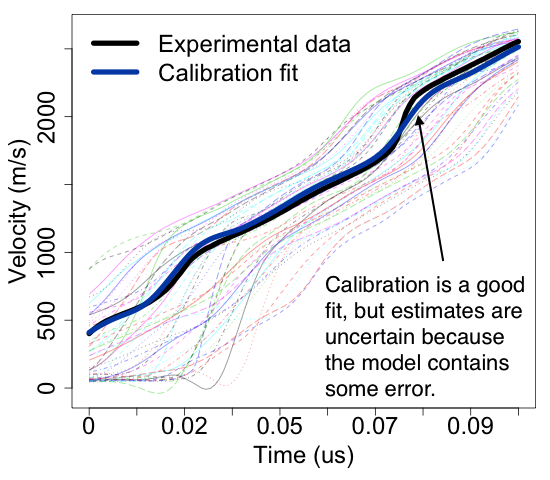}
\caption{Example of model discrepancy in a single experiment.}
\label{fig:expdisc}
\end{figure}

In this application, model discrepancy arises because there is no
combination of model inputs for which the computationally predicted
velocity curve can exactly match the experimental measurement (Figure
\ref{fig:expdisc}).  Whether this arises due to experimental
uncertainties that distort the velocity curve or due to actual
computational model error does not matter here for the mathematical
formulation of discrepancy. Furthermore, since the goal is quantifying
uncertainty in these physical parameters for the sake of extrapolative
prediction (i.e., prediction in contexts outside these experiments),
our method of Gibbs posteriors for model calibration is appropriate to
use here.  Since we have nine different experiments, with separate
functional discrepancies expected for each, it is also convenient to
calibrate each one separately and combine their respective posteriors
using our ensemble approach.

For the $j$\textsuperscript{th} experiment, we consider modeling the
output as $y(x_j) = \eta(x_j; \theta) + \delta(x_j) $, where
$\eta(x_{j}; \theta)$ is the computational model prediction of
velocity at $n_j=n=100$ time points $x_j$ with input values of the
calibration parameters $\theta$, and $\delta(x_j)$ is the discrepancy
between the model prediction and experimental measurement.  Failing to
account for model discrepancy would essentially result in the
uncertainty of the calibration parameters decreasing arbitrarily with
the number of points sampled from the functional output.

In the original analysis of these data, \cite{brown2018} apply both a
standard KOH model and a power-likelihood approach to account for
model discrepancy. Equivalently, we may say they implemented a Gibbs
posterior with the loss function being the Gaussian negative
log-likelihood,
\begin{align}\label{eq:loss-brown}
  l(y_j, \theta, \sigma) =
  \frac{n}{2} \log(2\pi\sigma^2) +
  \sum_{i=1}^{n} \frac{[y(x_{ji}) - \eta(x_{ji}; \theta)]^2}{2\sigma^2}.
\end{align}  
Although we previously made the comment that we can effectively assume
no error in the experimental measurements, the $\sigma^2$ parameter is
included in this loss function to quantify the scale of the
discrepancy, which is also potentially useful to compare across
experiments. To select the loss scales $w_j$ for the Gibbs posteriors
$p_w(\theta \mid y_j) \propto \exp(-w_j l(y_j, \theta, \sigma))
\pi(\theta)$, they first obtain a point estimate for the parameters
giving the best fitting output to the data,
$(\hat \theta_j, \hat \sigma_j) = \arg \min_{\theta, \sigma} l(y_j,
\theta, \sigma)$ and then select $w_{j}$ based on the effective sample
size (ESS) $n_{ej}$ of the empirical discrepancy
$\hat \delta(x_j) = y(x_j) - \eta(x_j; \hat \theta_j)$, so that
$\hat w_j^\text{ESS} = n_{ej} / n$.  To compute the full posterior for
$\theta$, they use the joint loss function, assuming independence
between experiments,
\begin{align}
  \label{eq:bh-joint-posterior}
  p(\theta \mid y) \propto \prod_{j=1}^9 \exp(- \hat w^\text{ESS}_j
  l(y_j, \theta, \sigma)) \pi(\theta).
\end{align} They conclude that the Gibbs
posterior generates more computationally stable results that are
consistent with prior knowledge about the material properties.

% \sw{Lauren: previously in this section you mention that we can ignore
%   error variance, but that is contradicted in this paragraph and in
%   the fact that the loss function \eqref{eq:loss-brown} has a $\sigma$
%   component.  So this section should be edited for clarity.  Also,
%   this paragraph feels extraneous anyway so maybe we should just cut
%   it. } Note that the loss function in Eq. \eqref{eq:loss-brown} can
% be extended to other likelihoods.  Further, the error-variance model
% can be extended to allow for residual overdispersion, i.e.,
% $\sigma(x) = \phi \sigma_m(x)$, where $\phi$ accounts for
% overdispersion in residuals beyond measurement uncertainty - a type of
% model misspecification).

Here we reanalyze the data making two changes.  First, we implement
the parametric bootstrap procedure of
Section~\ref{sec:parametric-bootstrap} to tune the choice of loss
scales $w_j$ for each experiment.  We use these to obtain posteriors
for each experiment independently.  Then, we use the WASP method of
ensemble calibration in Section~\ref{sec:ensemble-calibration} to
combine the posteriors for each experiment into one consensus
posterior.  For the sake of comparison, we will also calculate the
WASP consensus posterior using posteriors formed using loss scales
from the ESS method.  We use the same loss function
\eqref{eq:loss-brown} for the Gibbs posterior, and assign the weakly
informative priors $B'_0 \sim \mathcal{U}(2.9, 4.9)$ and
$\sigma^2 \sim \mathcal{IG}(0.01, 0.01)$.

We use the empirical Bayes variant of the parametric bootstrap
procedure, assuming that the discrepancy follows a mean-0 GP with
squared exponential covariance kernel (for simplicity we drop the $j$
subscript and explain the procedure for one experiment).  We first
calculate the empirical discrepancy term $\hat \delta(x)$ exactly as
done by \cite{brown2018}.  Then we estimate the hyperparameters for
the Gaussian process using maximum marginal likelihood on
$\hat \delta$ % and the estimated residual variance
% $\hat{\sigma}^2$ using maximum marginal likelihood on this empirical
% discrepancy,
to calculate an estimate of it covariance matrix $\hat \Sigma$.

One bootstrap dataset is calculated as follows.  First, take a draw
from the prior for $\theta$, $\theta^{(b)} \sim \pi(\theta)$, and
sample a bootstrap discrepancy term using
$\delta^{(b)} \sim \N(0, \hat
\Sigma)$. % , and a bootstrap sample for the
% error term using $\epsilon^{(b)} \sim \N(0, \hat{\sigma}^2\I)$
The bootstrap dataset is calculated by
$y^{(b)}(x)= \eta(x; \theta^{(b)}) + \delta^{(b)}(x)$.  We then form
equal-tailed Gibbs posterior credible intervals using $y^{(b)}$ for a
defined grid of possible loss scale values $\{w\}$, and check whether
these intervals bracket $\theta^{(b)}$.  Following this procedure for
$b=1,\ldots,B = 100$ gives a Monte Carlo estimate of frequentist
coverage for the Gibbs posterior with these values of $\{w\}$.  We
considered 35 values of $w$ on a log-scale for each experiment, and
then fit an interpolating spline to estimate coverage as a function of
$w$, with the inverse standard error as weights for fitting the
spline, and use a root finding algorithm to find which $w$ gives the
nominal frequentist coverage of $1-\alpha=0.9$.

\begin{figure}[!ht]
  \centering
  \includegraphics[width=0.85\textwidth]{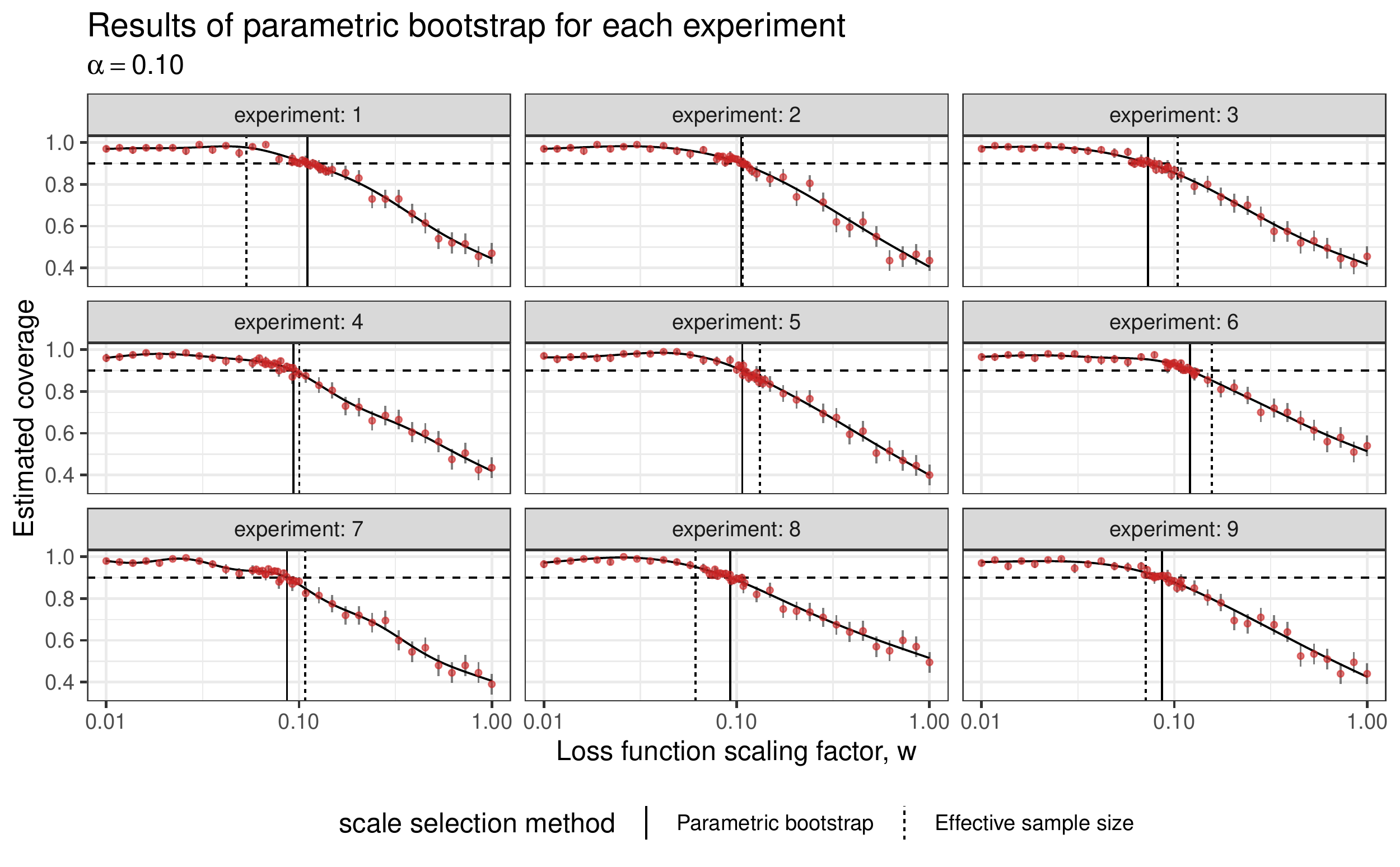}
  \caption{\label{fig:cover-plot-avg} Results of parametric bootstrap
    procedure for choosing the loss function selecting scaling factor.  
    % The estimated coverage on the $y$-axis is an average of the
    % coverage between the two parameters, $B_0$ and $B_0'$.
  }
\end{figure}

Figure~\ref{fig:cover-plot-avg} shows the results of the parametric
bootstrap procedure for choosing the Gibbs posterior loss scales
$w_{j}$ for each experiment $j=1,\ldots, 9$, and compares them to
those calculated from the residual ESS.  At times the two differ
widely, suggesting that the loss scales from ESS method could result
in poor frequentist metrics. In addition, the loss scales from the
parametric bootstrap tend to be closer to one another than those from
ESS.  

\begin{figure}[!ht]
  \centering
  \includegraphics[width=0.95\textwidth]{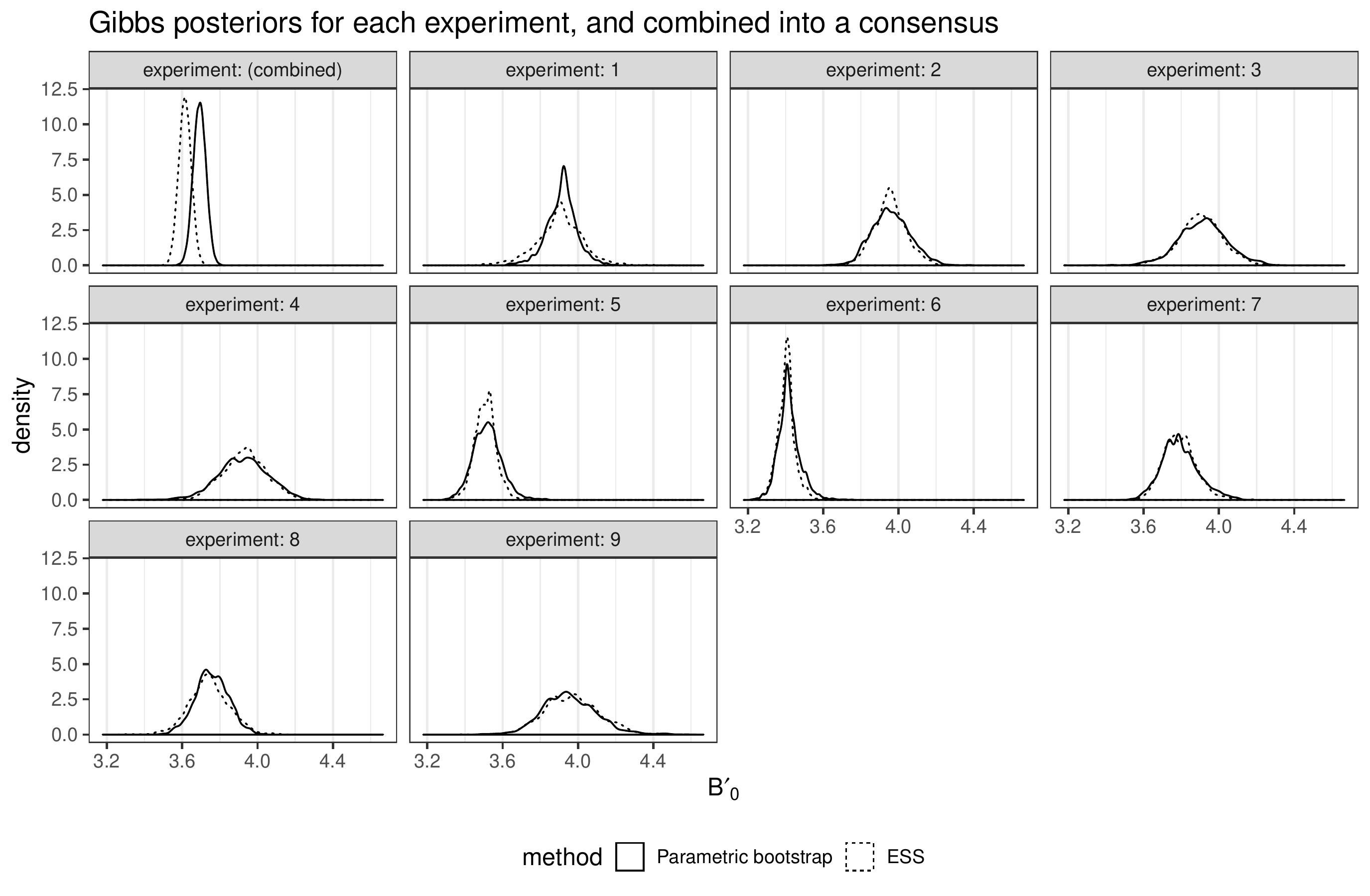}
  \caption{\label{fig:gibbs-combined} Gibbs posteriors for $B'_0$ from
    each experiment, and combined into a consensus posterior using the
    WASP ensemble calibration method.  We consider both methods of
    loss scale selection, our parametric bootstrap approach and the
    ESS method.  }
\end{figure}

Finally, Figure~\ref{fig:gibbs-combined} compares the
experiment-specific posteriors and the consensus posteriors resulting
from the two scaling factor selection methods for the parameter of
interest.  Because we combine inference across distinct experiments,
we scale the consensus variance by a factor of $1/9$, as explained in
Section~\ref{sec:comb-subs-post}. Both consensus posteriors have
similar variances, likely because the scaling factors from the two
methods are, on average, similar in magnitude.  However, since each
experiment is weighted differently between the two methods, the means
are quite far apart.  This difference is due to deviations in the
experiment-specific posterior variances; ESS has lower posterior
variance for experiments 5 and 6, which both have relatively smaller
posterior means, while the parametric bootstrap has lower posterior
variance for experiment 1, which has a relatively high posterior mean.
This explains the difference in consensus posterior means.  The online
supplement gives the loss scales and subset posteriors for each
experiment for both methods.

The consensus posterior distribution presented here differs from the
posterior in \cite{brown2018} due to differences in the estimation
procedure.  Specifically, the consensus posterior is based on a
weighted combination of experiment-specific posteriors, rather than
the joint posterior distribution in Eq.~\eqref{eq:bh-joint-posterior}.
The point-estimate of the parameter is an inverse-variance weighted
average of the experiment specific point estimates, and, from
Figure~\ref{fig:gibbs-combined}, we can see that two experiments
(experiments 5 and 6) have small variances and also result in much
lower estimates of $B'_0$ than the remaining experiments.  Hence,
these experiments have substantial influence over the consensus
posterior.

We consider this property at once to have both strengths and
drawbacks.  One possible disadvantage is that our method may be
sensitive to outlying sets of data.  However, the ability to examine
experiment-specific posteriors and identify influential sets of data,
such as experiments 5 and 6, is a key advantage of the consensus
posterior approach.  The experiment specific posteriors for $B'_0$ are
shown in Figure~\ref{fig:gibbs-combined}.  We can consider whether the
experiment-specific estimates are consistent with scientific knowledge
or appear anomalous, which informs where model discrepancy may exist
within the support of the data.  This information may help identify
areas where the model could be improved and/or where we might place
less faith in the physical parameter estimates.

% The resulting posteriors for both methods differ from those from the
% original results from \cite{brown2018}; the posterior means
% for $B'_0$ are slightly lower, and the posterior variance is also lower
% than originally reported.  We believe these disparities to arise from
% using a different emulator for the computer model for
% $\eta(x; \theta)$.  We used an emulator based on multivariate adaptive
% regression splines (MARS) \citep{friedman1991}, rather than Gaussian
% process regression used by the original paper.

%%% Local Variables:
%%% mode: latex
%%% TeX-master: "../Draft"
%%% End:

\section{Discussion}
\label{sec:discussion}

When computational models are used for extrapolative prediciton, there
is a need to have well-defined calibration parameters and use
statistical methods which make inference more robust to model
misspecification.  Estimating a discrepancy function is not necessary
in physical parameter estimation is the primary focus.  With this
motivation, we have introduced Gibbs posteriors for Bayesian model
calibration when the goal is physical parameter estimation for
extrapolative prediction.  % In contrast to the existing standard
% Bayesian approach, and in connection to literature casting calibration
% as an optimization problem, the target of inference is the parameter
% minimizing an expected loss.
This generalizes previous power-likelihood approaches to model
calibration.

The loss scale in the Gibbs posterior can be tuned to maintain nominal
frequentist coverage under assumptions of the discrepancy form, and
coverage is approximated using our bootstrap procedure.  This
procedure shares conceptual similarities to the pre-posterior method
from \cite{arendt2016preposterior} to determine how identifiable a
calibration parameter is by repeatedly sampling discrepancy functions
and evaluating variability in the calibration parameters.  It is also
similar to the bootstrap procedure used by \cite{wong2017frequentist}
to estimate the sampling distribution of the frequentist point
estimate in Eq.~\eqref{eq:point-wong}; however, the underlying goal in our
case is much different, as our bootstrap procedure is designed to
calibrate the width of credible intervals from the Gibbs posterior.

There are several advantages of using the Gibbs posterior for physical
parameter estimation over the existing Bayesian approach.  First, the
target of inference is well defined by the specified loss function, so
we largely avoid the typical issues of unidentifiability.  Second, the
sampling scheme for Gibbs posterior is more computationally stable and
less expensive than that of the current standard Bayesian method, as
now we do not need to calculate matrix inversions necessary when
sampling the posterior for the Gaussian process discrepancy
term. Finally, our approach is highly modular; it allows for various
choices of loss function to define the target of inference, and
provides the analyst a large degree of freedom in specifying forms of
model discrepancy to inform the choice of loss scale.

Additionally, we present a method of ensemble calibration, making the
case that it is often both advantageous to calibrate separately on
data subsets, and then combine their respective posteriors into a
consensus posterior.  An implicit assumption here is independence
between subsets, and this may not always be the case.  For example, in
our application in Section~\ref{sec:dataapp}, the discrepancy largely
depends on the experimental conditions for each data subset, and these
are likely correlated.  This could be handled, for example, by
incorporating dependence when combining partitions.  To our knowledge,
how to do this remains an open question.

Because our approach is intentionally general, there is much promise
for future work.  We mainly focused on using the $\ell_2$ loss to
define the inferential target, and so it may be of interest to find
applications where other loss functions are suitable.  It would also
be informative to find an application where the calibrated Gibbs
posterior can be directly validated against external data or known
parameter values.  There are other potential bootstrapping schemes for
approximating frequentist coverage to be explored, and likely more
efficient means for tuning the loss scale, such as an iterative method
rather than a grid search.

\long\def\acks#1{\vskip 0.3in\noindent{\large\bf Acknowledgments}\vskip 0.2in
\noindent #1}

\acks{We would like to thank Professor Peter M\"uller at the
  University of Texas at Austin for helpful discussion, and Justin
  Brown at Sandia National Laboratories for allowing us to use  his data for the
  application.

  This work was supported by a Sandia National Laboratories Laboratory
  Directed Research and Development (LDRD) grant. Sandia National
  Laboratories is a multimission laboratory managed and operated by
  National Technology and Engineering Solutions of Sandia, LLC., a
  wholly owned subsidiary of Honeywell International, Inc., for the
  U.S. Department of Energy's National Nuclear Security Administration
  under contract DE-NA0003525. Approved for unlimited release,
  SAND2019-10771 R. This paper describes objective technical results
  and analysis. Any subjective views or opinions that might be
  expressed in the paper do not necessarily represent the views of the
  U.S. Department of Energy or the United States Government.}

%%% Local Variables:
%%% mode: latex
%%% TeX-master: "../Draft"
%%% End:

% -------------------------------------------------------------------------
% Appendices

\appendix

\section{Alternative bootstrapping techniques}

\subsection{Non-parametric bootstrapping}
\label{sec:np}

To implement the bootstrap in Section 3, a non-parametric bootstrap
could also be applied for weight selection, as in
\cite{syring2017calibrating}, where resampling with replacement is
used to approximate the uncertain data generating process $F_0^*$.
The key methodological question in the calibration setting is how to
resample the data to accurately represent prior uncertainty associated
with model discrepancy.  Standard approaches, such as simply
resampling observations with replacement, must be adjusted to account
for the fact that observations are not independent under model
discrepancy.  One way to address this lack of independence is using a
block bootstrap, partitioning the data into approximately independent,
non-overlapping blocks and sampling the blocks with replacement
\citep{kreiss2011bootstrap}.  Blocks should be selected such that
model discrepancy is approximately constant over the blocks.

The validity of this nonparametric method is highly dependent on our
ability to partition the data into independent blocks.  Selecting
blocks that are too small will result in residual dependence between
blocks and subsequent underestimation of uncertainty in
$\theta$. Choosing blocks that are too large will result in an
inability to accurately approximate the sampling distribution of $\theta$ under uncertainty in the data generating mechanism,
again resulting in underestimation of uncertainty.  With prior
information about where the model discrepancy may vary over $\{x\}$,
the blocks can be selected using subject matter expertise.  In the
absence of such knowledge, the block size can be selected based on the
autocorrelation time, assuming a stationary, mean 0 discrepancy
function, resulting in equally-sized blocks.  If the blocks contain
different numbers of observations, the analyst must adapt the
bootstrap appropriately \citep{sherman1997comparison}.
If the model is systematically biased such that the expected
value of the model discrepancy is non-zero, inferences on $\theta$ will be biased unless the loss function and resampling procedure are appropriately updated using subject matter knowledge about the expected value of the discrepancy.

This procedure is very similar to the bootstrap procedure from
\cite{syring2017calibrating}, but adapted to the autocorrelated nature
of the data in the model calibration problem.  More efficient
implementations of this algorithm are possible by searching over
$\{ w \}$ in an informed way.  When $\theta$ is multivariate, a joint
credible region must be constructed.

Practical implementation of the nonparametric bootstrap can be impeded
by the following disadvantages: (1) the data must be partitioned into
independent blocks, which may be challenging in practice; and (2) the bootstrap will be inaccurate when the number of partitions is small, but the number
of partitions is determined by the nature of the model discrepancy and
cannot be arbitrarily increased.

%%% Local Variables:
%%% mode: latex
%%% TeX-master: "../Draft"
%%% End:

\subsection{Alternative parametric bootstrap algorithm}
\label{sec:appendix_pboot}

The bootstrap procedure below more closely mimics a classic
frequentist parametric bootstrap, as described in, for instance,
\cite{syring2017calibrating}.  The procedure works as follows:
\begin{enumerate}[(i)]
\item Given a value of $w$, construct credible intervals
  $C_{\alpha, w}(\theta \mid y)$.  Generate a point estimate of $\theta$, denoted $\hat{\theta}$, from the fitted model.
\item Create bootstrap sample data.  For $b=1,\ldots,B$, Draw
  components, then the data is 
  \begin{align*}
    \delta^{(b)} &\sim \pi(\delta \mid {\lambda}) \\
    \epsilon^{(b)} &\sim \pi(\epsilon \mid \tau)  \\
    y^{(b)}(x)&= \eta(x; \hat{\theta}) + \delta^{(b)}(x) + \epsilon^{(b)}(x)
  \end{align*}
\item Construct a MAP estimate of $\theta$ from the model with loss
  scale $w$ for each bootstrap sample, denoted $\hat{\theta}^{(b)}$.
\item The frequentist coverage is estimated by
  \begin{align*}
    \hat{c}_\alpha(w; F^\star_0)
    &= B^{-1}\sum_{b=1}^B
      \mathbf{1}(\hat{\theta}^{(b)} \in C_{\alpha, w}(\theta \mid y))
  \end{align*}
\item Choose the value of $w$ for which
  $\hat{c}_\alpha(w; F^\star_0) \approx 1 - \alpha$, by using a
  defined grid of values or by using a stochastic approximation
  \citep{syring2017calibrating}.
\end{enumerate}

Once the optimal scale $\hat w$ is selected, we can form the calibrated
Gibbs posterior using the experimental data.  

%%% Local Variables:
%%% mode: latex
%%% TeX-master: "../Draft"
%%% End:

\subsection{Comparison with cross validation}
\label{sec:cv}

Cross validation is a commonly used approach to selecting $w$, and therefore we briefly discuss how cross-validation could be applied in the calibration application.
Key decisions in implementing cross validation are: (1) how to partition the data, and (2) how to select cross-validation metrics for model selection:
\begin{itemize}
\item \textbf{Partitioning the data.}  As with the nonparametric bootstrapping method (Section \ref{sec:np}), when constructing cross-validation subsets, correlation between observations must be considered, because calibration data are not \textit{iid} observations in the presence of model discrepancy.  

\item \textbf{Metrics.} Cross-validation considers the predictive
  ability of the model outcome $y$ to the left-out test set.  We again
  are more concerned with uncertainty quantification than predictive
  accuracy, so the metric for selecting $w$ should be frequentist
  coverage of the model output based on the posterior predictive
  distribution.

% Suppose the data are partitioned into $k=1,...,K$ subsets.   For test dataset $k$, the data are trained on the remaining $K-1$ parts and tested on subset $k$.  Let $\tilde{y}_k(x)|y_{(-k)}$ denote the test-data posterior predictive distribution of $y_k$ given a posterior for $\theta$ estimated used only the training data.  Let $(\tilde{y}^L_k, \tilde{y}^U_k)$ denote a 95\% posterior predictive interval for $y_k$.  Then, we can select $w$ based on:
% \begin{eqnarray}
% \hat{w} &=& \mathrm{argmax}_w E_k \left( P[y_k \in (\tilde{y}^L_k, \tilde{y}^U_k)] \right ) \geq (1-\alpha) 
% \end{eqnarray}
% The outer expectation is taken over the different test sets $k$.  
\end{itemize}

\head{Challenges in implementation.} Challenges in implementing
cross-validation are similar to the nonparametric bootstrapping
challenges (Section \ref{sec:np}).  Specifically, an important
assumption behind this cross-validation procedure is that the
partitions are $iid$ samples.  To achieve approximate independence
between the $K$ partitions, the number of selected partitions may have
to be small.  When $K$ is small, then the cross-validation metrics may
be highly uncertain (particularly coverage).  Additionally, if some
subset of the calibration parameters vary over $\{x\}$, then
implementing cross validation is not straightforward.  Section 5 of
the main text includes such an example.

%%% Local Variables:
%%% mode: latex
%%% TeX-master: "../Draft"
%%% End:

\section{Simulation study for parametric bootstrap}
\label{sec:simulation-study}

We conduct a simulation study to test whether the parametric
bootstrapping weight selection procedure achieves approximately
nominal frequentist coverage for a single dynamic material properties
experiment.

\begin{figure}[!ht]
  \centering
  \includegraphics[width=.3\textwidth]{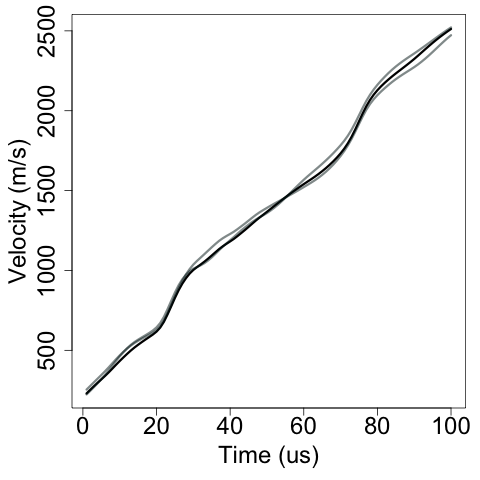}
  \includegraphics[width=.3\textwidth]{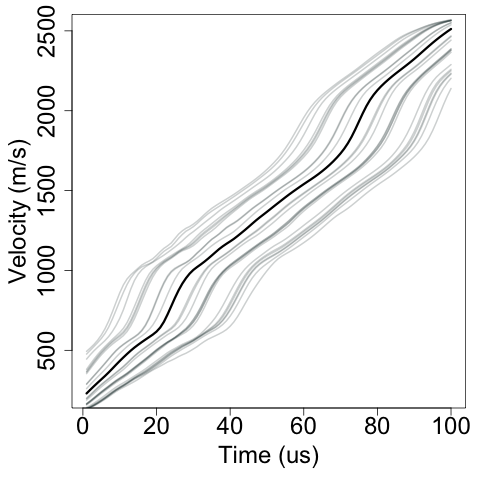}
  \caption{Left: ``True'' velocity curve (black) and two simulated
    velocity curves with model discrepancy with high autcorrelation time
    (grey).  Right: ``True'' velocity curve at the correct value of the
    calibration parameter (black) and simulated velocity curves at
    different draws from the prior distribution on the calibration
    parameter.}
  \label{fig:simdata}
\end{figure}

\textbf{Data generating mechanism.} To define the data generating
model, we fix the ``true'' calibration parameters to their prior means
for a single experiment.  We assume the true model discrepancy is a
mean 0 Gaussian process, with a squared exponential kernel.  Example
simulated datasets are shown in Figure \ref{fig:simdata}, where the
magnitude of the model discrepancy can be seen relative to the prior
uncertainty in the calibration parameters.  We use two different range
parameters to determine how the shape of the discrepancy impacts the
results.  Specifically, the range parameter was selected such that the
autocorrelation time of the Gaussian process is approximately either
1/5 or 1/10 the support of the experiment time $t \in \mathcal{T}$.
We refer to these settings as high versus low autocorrelation time.

\textbf{Fitted model.}  For simplicity, we only estimate a single
calibration parameter, the pressure derivative $B_0'$.  The remaining
calibration parameters are fixed to their true values.  To estimate
the calibration parameter $B_0'$ and its uncertainty, we use the
Gaussian negative log-likelihood as our loss function:
\begin{align*}
  l(y, B_0', \sigma) =
  \frac{T}{2} \log(2\pi\sigma^2) +
  \sum_{t=1}^{T} \frac{[y_t - \eta_t(B_0')]^2}{2\sigma^2}
\end{align*}  
for $t=1,...,T=100$.  We use the same weakly informative prior on the
calibration parameters as in Section~5 of the main text.  A
non-informative inverse gamma prior was placed on the variance
parameter $\sigma$.  A flexible generalize additive model was used to
develop a computationally inexpensive surrogate model for
$\eta_t(B_0')$.  The accuracy of the surrogate is sufficiently high
such that we anticipate that contribution of surrogate error to the
simulation results will be negligible.

\textbf{Methods} We compare weight selection based on the parametric
bootstrap to the previously proposed effective sample size method
described in \cite{brown2018}.  These approaches depend on tuning
parameters.  Specifically, the effective sample size method depends on
the estimated autocorrelation time; and the parametric bootstrap
depends on the assumed form of the model discrepancy.  For each
method, we run the simulation study in two cases: (1) fixing the
tuning parameters to their true values and (2) estimating the tuning
parameters.

Using the effective sample size method, the weight was selected based
on the autocorrelation time of the model discrepancy term.  To
estimate the autocorrleation time of the discrepancy, the
computational prediction of the velocity curve at the maximum a
posteriori (MAP) estimate of the calibration parameters was subtracted
from the experimental data to provide a MAP estimate of the model
discrepancy.  The autocorrelation time was then estimated on the
estimated model discrepancy using the formula described in
\cite{brown2018}.

In the parametric bootstrap, the weight was selected by assuming the
model discrepancy follows a mean-zero Gaussian process (note that,
given the data generating mechanism, this assumption is correct).
When estimating the correlation function of the GP, we correctly
assumed a Gaussian correlation functions and used a maximum likelihood
estimate of the range parameter based on the estimated model
discrepancy.  To speed up the simulations, we used a maximum
likelihood approximation to the full Bayesian model when conducting
the grid search to find an optimal $w$.

For the bootstrap models, a grid search over different weight values
was implemented, using the grid of weights
$\{.01, ..., .09, 0.10..., 1\}$.  The final weight was selected as the
smallest weight for which the model's 90\% credible interval from the
power-likelihood posterior achieved at least 90\% coverage over the
bootstrap resamples.  To estimate coverage, we ran 100 Monte Carlo
simulations (due to the computational expense of the procedure).

\textbf{Results.}  Both the parametric bootstrap and effective sample size methods produce estimate of $w$ that are reasonably close to the true value (represented by the autocorrelation time).  Further, coverage is close to the target 90\% nominal coverage.

% latex table generated in R 3.5.2 by xtable 1.8-3 package
% Tue Feb 26 13:27:13 2019
\begin{table}[ht]
\centering
\begin{tabular}{l|c|rr|rr}
 & Autocorrelation & \multicolumn{2}{c|}{Fixed} & \multicolumn{2}{c}{Estimated} \\ 
 & time & E(w) & Coverage & E(w) & Coverage \\ 
  \hline
%Non-parametric bootstrap & 0.14 & 0.76 & 0.15 & 0.79 \\ 
  Parametric bootstrap & 0.1 & 0.08 & 0.95 & 0.10 & 0.92 \\ 
  Effective sample size & 0.1 & 0.10 & 0.94 & 0.13 & 0.91 \\ 
   Parametric bootstrap & 0.2 & 0.03 & 0.99 & 0.05 & 0.89 \\ 
   Effective sample size & 0.2 & 0.05 & 0.93 & 0.08 & 0.86 \\ 
  \end{tabular}
\caption{Frequentist coverage for two autocorrelation times, where autocorrelation time is parameterized in terms of the fraction of the support at which the autocorrelation is approximately 0 (and should correspond approximately to $w$).  
       Nominal coverage is 90\%.\label{tab:simresults1}} 
\end{table}

%%% Local Variables:
%%% mode: latex
%%% TeX-master: "../Draft"
%%% End:

\section{Subset posterior calculation}
\label{sec:wass}

In this section we outline the algorithm for combining subset
posteriors into a consensus posterior with WASP that was introduced in
Section~4 of the main text.

When calibrating each experiment separately to obtain the constituent
posteriors,
\begin{eqnarray} p(\theta \mid y_k) \propto \exp(-w_k
l(y_k \mid \theta))\pi(\theta), \quad k=1,\ldots,K \label{eq:obayes2}
\end{eqnarray} 
we must be cautious not to use the prior $K$ different times, as this
would allow the prior to have undue influence on the end results.  To
obtain the correct prior influence for the consensus posterior using
WASP, the likelihood is reweighted using a ``stochastic approximation
trick,'' \citep{pmlr-v38-srivastava15}. Specifically, for $K$ subsets
of equal size, each subset likelihood should be raised to the power
$K$. Essentially we are adding $K-1$ copies of each observation in
each subset, making the implied sample size for each subset
commensurate with the full data sample
size. %Each subset posterior a `noisy' image of the full data posterior, and it insures the correct scale for the consensus posterior and its credible intervals. \lh{to clarify here, when you are initially estimating subset posteriors, you need to raise the likelihood to the power of $K$.  in our case, we will first find $w$, then estimate the subset posterior by raising the likelihood to $Kw$.}

%The WASP method does not require the same independence structure between subsets as CMC, and it is far more robust to the subsetting process.

To combine subset posteriors, we use a fixed-point algorithm for
calculating quadratic Wasserstein barycenters by
\citep{alvarezestebanetal2016}.  The consensus location parameter is
the inverse covariance-weighted average of the subset location
parameters. % \lh{this seems really counterintuitive to me... wouldn't
  % you want to use a weighted average, weighting distributions with
  % lower variance more heavily than distributions with higher variance;
  % i.e., why not inverse variance weight??}
The consensus covariance matrix is calculated by finding the
covariance matrix of the barycenter for the consensus distribution's
covariance.  For detailed derivation, see
\cite{alvarezestebanetal2016}.

The algorithm proceeds as follows.
\begin{description}
	\item[(1)] Divide data $K$ subsets, calculate each subset
posterior $p(\theta|y_k)\propto\exp(-Kw_kl(y_k|\theta))\pi(\theta)$
	\item[(2)] Obtain covariance matrices $\Sigma_k$ and means
$\mu_k$ for each subset posterior
	\item[(3)] Select a starting $S_0$ (e.g. $S_0:=I$),
$\varepsilon>0$ and $i:=1$, where $S_i$ is the `guess' for the
barycenter covariance matrix on iteration $i$.
	\item[(3a)] Calculate $S_i$ , defined as:
$$S_{i}=S_{i-1}^{-1/2}\left(\sum_{k=1}^{K}\tfrac{1}{K}(S_{i-1}^{1/2}\sigma_k{}S_{i-1}^{1/2}S^{1/2}_{i-1})^{1/2}\right)^2S_{i-1}^{-1/2}$$
	\item[(3b)] if $\Vert{}S_i-S_{i-1}\Vert>\epsilon$ (for some
appropriate matrix norm), set $i:=i+1$ and $\mathrm{goto}$ 3a, else
set $\bar{S}=S_i$ and continue
	\item[(4)] Set $\bar{\mu}=(1/K)\sum_{k}\Sigma_k^{-1}\mu_k$
\end{description}
The output is location $\bar{\mu}$ and a covariance $\bar{S}$
describing the consensus posterior that is in the same location-scale
family.  The algorithm tends to converge very rapidly.

When all distributions being `averaged' in a Wasserstein barycenter are in the same location-scale family, the algorithm converges quickly and produces an exact barycenter, also in the same location-scale family. This is well-suited to our scenario where misspecification is addressed by power likelihoods in a Gaussian setting. Raising each likelihood to a fractional power can be seen as rescaling each posterior distribution to account for misspecification and discrepancy.
 %Each experiment can be calibrated separately, reflecting
%uncertainty particular to that experiment, and combined into a
%consensus posterior aggregating this information. 

The fixed-point algorithm that we use to combine subset posteriors is a more restrictive, but more accurate and cost efficient method than the full WASP procedure described in \citep{pmlr-v38-srivastava15}.   Computationally, the full WASP procedure is implemented by first sampling subset posteriors. The approximate barycenter of these subset posterior samples is obtained through linear programming. A consensus distribution, supported on a subset of all the subset posterior points sampled, is found by minimizing the
Wasserstein distance from each subset posterior to the
consensus.   A limiting factor for the WASP is computational feasibility; when subsets are numerous and each subset posterior has to be sampled many times, the linear program may be too costly for implementation. 

An alternative popular technique for combining subset posteriors is
consensus Monte Carlo (CMC).  However, CMC has several properties that
make it less preferable in calibration settings.  Unlike WASP, the
prior is ``flattened'' for each subsample by raising it to the power
$1/K$; this avoids placing too much weight on the prior in the
consensus posterior. Inference is conducted and subset posteriors
$\Pi_i$, $i=1,\ldots,K$, are weighted and combined into a consensus
posterior $\hat{\Pi}$.  However, in calibration problems, the prior
distribution is often critical to parameter identifiability.
Down-weighting the prior may produce non-sensical inferences in some
subsets. It can also eliminate conjugacy of a Bayesian model. When
each subset posterior is normal, CMC produces exact draws from the
full data posterior. While CMC is robust to some deviation from
Gaussianity, its accuracy is dependent on how far the true
data-generating mechanism is from normality as well as the sample
size. For a full account of CMC, see \cite{scott2016bayes}.

%%% Local Variables:
%%% mode: latex
%%% TeX-master: "../Draft"
%%% End:

\section{Additional figures}
\label{sec:extra-plots}

\begin{figure}[!ht]
  \centering
  \includegraphics[width=0.7\textwidth]{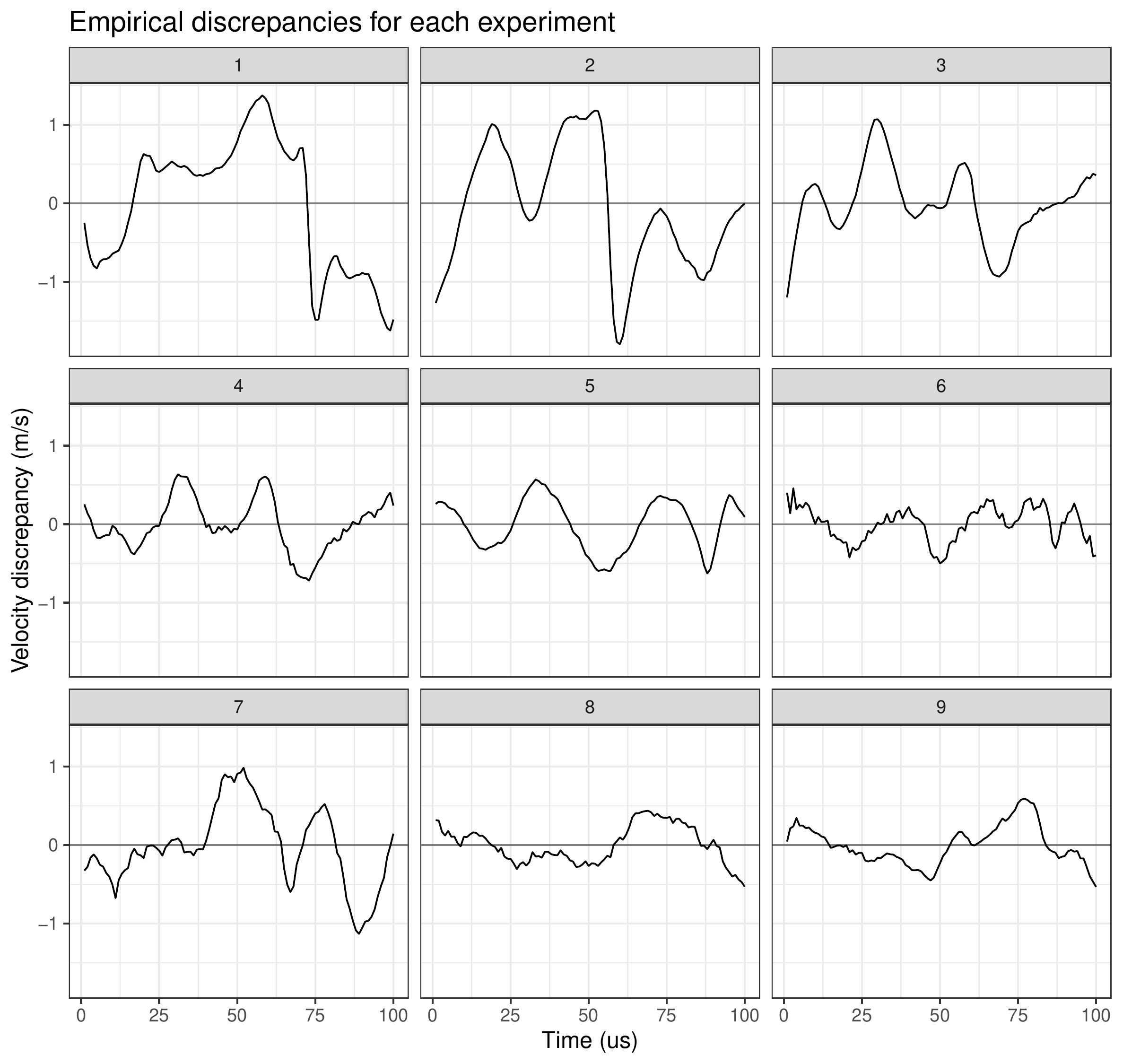}
  \caption{\label{fig:disc-plot} Empirical discrepancy functions for all experiments.}
\end{figure}

See Figure~\ref{fig:disc-plot}. % , \ref{fig:cover-plot-grid}, and
% \ref{fig:gibbs-posteriors-wrap}

\singlespacing
\bibliography{Draft}
\end{document}